\documentclass{aa}
\usepackage{graphicx}
\usepackage{txfonts}
\usepackage{natbib}
\bibpunct{(}{)}{;}{a}{}{,}

\usepackage{lscape}
\usepackage{longtable}
\usepackage{amssymb}
\usepackage{xfrac}
\usepackage{color}

\begin{document}

% Version 1
%   \title{High-resolution radio continuum survey towards luminous YSOs}
%   \subtitle{...}
%
% Version 2
%   \title{Outflow structure within 1000\,au of high-mass YSOs}
%   \subtitle{II. High-resolution radio continuum survey of H$_2$O maser sources}
%
% Version 3
   \title{Protostellar Outflows at the EarliesT Stages (POETS)}
%   \subtitle{II. High-resolution VLA continuum survey of H$_2$O maser sources}
  \subtitle{I. Radio thermal jets at high resolution nearby H$_2$O maser sources}

   \author{A. Sanna\inst{1,2} \and L. Moscadelli\inst{3} \and C. Goddi\inst{4} \and  V. Krishnan\inst{3} \and  F. Massi\inst{3}}
% \and R. Cesaroni\inst{2} \and \'{A}. S\'{a}nchez-Monge\inst{4}
            
%\fnmsep\thanks{Just to show the usage of the elements in the author field}

   \institute{Max-Planck-Institut f\"{u}r Radioastronomie, Auf dem H\"{u}gel 69, 53121 Bonn, Germany \\
   \email{asanna@mpifr-bonn.mpg.de}
   \and INAF, Osservatorio Astronomico di Cagliari, via della Scienza 5, 09047 Selargius (CA), Italy   
   \and INAF, Osservatorio Astrofisico di Arcetri, Largo E. Fermi 5, 50125 Firenze, Italy
   \and Department of Astrophysics/IMAPP, Radboud University Nijmegen, PO Box 9010, NL-6500 GL Nijmegen, the Netherlands
    }
   \date{Received 5 June 2018 / Accepted 16 July 2018}

%   \and I. Physikalisches Institut, Universit\"{a}t zu K\"{o}ln, Z\"{u}lpicher Str. 77, 50937 K\"{o}ln, Germany

% \abstract{}{}{}{}{}
% 5 {} token are mandatory

  \abstract
  % context heading (optional)
  % {} leave it empty if necessary
   {Weak and compact radio continuum and H$_2$O masers are preferred tracers of the outflow activity nearby very young stars.}
  % aims heading (mandatory)
   {We want to image the centimeter free-free continuum emission, in the range 1--7\,cm (26--4\,GHz), which arises in the inner few
   1000\,au from those young stars also associated with bright H$_2$O masers. We want to study the radio continuum properties in
   combination with the H$_2$O maser kinematics, in order to eventually quantify the outflow energetics powered by single young stars.}
  % methods heading (mandatory)
   {We made use of the Karl G. Jansky Very Large Array (VLA) in the B configuration at K band, and in the A configuration at both Ku
   and C bands, in order to image the radio continuum emission towards 25 H$_2$O maser sites with an angular resolution and thermal
   rms of the order of $0\farcs1$ and 10\,$\mu$Jy\,beam$^{-1}$, respectively. These targets add to our pilot study of 11 maser
   sites presented in \citet{Moscadelli2016}. The sample of H$_2$O maser sites was selected among those regions having an accurate
   distance measurement, obtained through maser trigonometric parallaxes, and H$_2$O maser luminosities in excess of 10$^{-6}$\,L$_{\odot}$.}
  % results heading (mandatory)
   {We present high-resolution radio continuum images of 33 sources belonging to 25 star-forming regions. In each region, we detect
   radio continuum emission within a few 1000\,au of the H$_2$O masers' position; 50\% of the radio continuum sources are associated with
   bolometric luminosities exceeding 5\,$\times$\,10$^{3}$\,L$_{\odot}$, including W33A and G240.32$+$0.07. We provide a detailed spectral
   index analysis for each radio continuum source, based on the integrated fluxes at each frequency, and produce spectral index maps with the
   multi-frequency-synthesis deconvolution algorithm of CASA. The radio continuum emission traces thermal bremsstrahlung in (proto)stellar
   winds and jets, with flux densities at 22\,GHz below 3\,mJy, and spectral index values between $-0.1$ and 1.3. We prove a strong
   correlation ($r$\,$>$\,0.8) between the radio continuum luminosity (L$_{\rm rad}$) and the H$_2$O maser luminosity (L$_{\rm H_2O}$)
   of  (L$_{\rm 8GHz}/$mJy\,kpc$^2)$\,$=$\,10$^{3.8}$\,$\times$\,(L$_{\rm H_2O}/$\,L$_{\odot})^{0.74}$. This power-law relation is
   similar to that between the radio continuum and bolometric luminosities, which confirms earlier studies. Since H$_2$O masers are excited through
   shocks driven by (proto)stellar winds and jets, these results provide support to the idea that the radio continuum emission around young stars 
   is dominated by shock-ionization, and this holds over several orders of magnitude of stellar luminosites ($1-10^5$\,L$_{\odot}$).}
  % conclusions heading (optional), leave it empty if necessary
  {} 


  \titlerunning{High-resolution VLA continuum survey of H$_2$O maser sources}
   \maketitle
%
%________________________________________________________________

\section{Introduction}

Outflow activity is a proxy for ongoing star formation. A number of outflow studies have established a correlation between the
bolometric luminosity (L$_{\rm bol}$) of a star-forming region and the integrated (molecular) outflow properties, such as the
mechanical force and momentum released into the ambient gas, which, at parsec scales, quantify the overall contribution of
an ensemble of outflows \citep[e.g.,][and references therein]{Arce2007,Frank2014}. This relationship holds over six orders
of magnitude of L$_{\rm bol}$ (e.g., \citealt{Beuther2002}, their Fig.\,4; \citealt{Maud2015}, their Fig.\,7), and it is interpreted
as evidence for a single outflow mechanism which scales with the stellar luminosity, and the outflows' motion being momentum
driven. However, on scales of a few 1000\,au, representative of individual young stars, there are poor statistics on outflow
properties for stellar luminosities exceeding 10$^{3}$\,L$_{\odot}$.
%which, at distances of several kilo-parsec  from the Sun, require angular resolutions of the order of a tenth of arcseccond (and smaller).  

For the purpose of studying the dynamical properties of the outflow emission in the vicinity of luminous young stars, we have started
the ``Protostellar Outflows at the EarliesT Stages'' (POETS) survey. The target sample has been selected with the idea of combining
the kinematic information of outflowing gas, inferred from the H$_2$O maser emission, with the information of ejected mass, inferred
from the H~{\sc ii}, free-free, continuum emission (e.g., \citealt{Moscadelli2016,Sanna2016}).

On the one hand, H$_2$O maser emission traces shocked gas propagating in dense regions ($n_{\rm H_2}$\,$>$\,10$^6$\,cm$^{-3}$)
at velocities between 10 and 200\,km\,s$^{-1}$ \citep[e.g.,][]{Hollenbach2013}: these properties make H$_2$O masers 
signposts of (proto)stellar outflows within a few 1000\,au from their driving source. Maser spots, namely ``cloudlets''
of the order of a few au in size, are ideal test particles to measure the local three-dimensional motion of gas shocked 
where stellar winds and jets impact ambient gas \citep[e.g.,][]{Torrelles2003,Goddi2006b,Sanna2012,Moscadelli2013,Burns2016,Hunter2018}. 

On the other hand, thermal (bremsstrahlung) continuum emission, with flux densities lower than a few mJy, and spectral index values ($\alpha$)
at centimeter wavelengths between $-0.1$ and below 2, traces the ionized gas component of stellar winds and jets 
\citep{Panagia1975,Reynolds1986,Anglada1996,Anglada1998}. Ionization is caused by shocks around young stars with spectral types
later than B, which emit negligible Lyman flux \citep{Curiel1987,Curiel1989}. In these sources, the centimeter continuum luminosity scales with
the stellar luminosity as a power law of index 0.6, approximately \citep{Anglada1995,Anglada2015}.
Extrapolating this law to young stars of spectral types B and earlier, one expects, for instance, a continuum flux of approximately 1\,mJy 
at 8\,GHz, for a B1 zero-age-main-sequence (ZAMS) star at a distance of 1\,kpc from the Sun.                                                    % used eq. 3.1 of Anglada+2015 
For comparison, a homogeneous, optically thin H~{\sc ii} region, that is excited by the Lyman photons of a B1 ZAMS star, emits a continuum
flux more than two orders of magnitude higher ($\approx$\,0.2\,Jy), at the same frequency and distance.                                   % used eq. 2 of Guzman+2012
It follows that, ionized stellar winds and jets can be detected prior to the ultra-compact (UC) H~{\sc ii} region phase, which
implies typical lifetimes older than 10$^4$\,yr (e.g., \citealt{Churchwell1999,Churchwell2002,Hoare2007}).                

Recently, two distinct surveys have searched for ionized stellar winds and jets in regions with bolometric luminosities typically exceeding 
10$^{3}$\,L$_{\odot}$ \citep{Rosero2016,Purser2016}. \citeauthor{Rosero2016} made use of the Karl G. Jansky Very Large Array (VLA)
at 6 and 22\,GHz to survey 58 star-forming regions at different stages of evolution. They detected ionized jet emission in about half the sample,
with typical values of angular resolution and sensitivity of 0.4--0.3\,arcsec and 5--10\,$\mu$Jy\,beam$^{-1}$, respectively. \citeauthor{Purser2016}
made use of the Australia Telescope Compact Array (ATCA) at 5, 9, 17, and 22\,GHz to survey 49 star-forming regions selected from the RMS
survey. They detected ionized jet emission in 26 distinct sites (12 of which are candidates), with typical values of angular resolution and
sensitivity ranging between 2--0.5\,arcsec and 17--106\,$\mu$Jy\,beam$^{-1}$, respectively. In particular, \citeauthor{Purser2016} reported 
non-thermal knots in a subset of jets (10), which resemble the prototypical magnetized jet in HH\,80--81 \citep{Carrasco2010,RodriguezKamenetzky2017}.

In order to provide a homogeneous sample of stars at an early stage of evolution, we limited the sample to rich H$_2$O maser
sites not associated with UCH~{\sc ii} regions (where the continuum emission exceeds 10,000\,au in size). Among the H$_2$O maser sites
satisfying this requirement, we selected those targets whose distances were accurately determined by trigonometric parallax measurements,
and where the maser positions are known at milli-arcsecond resolution \citep{Reid2014}.
The POETS sample of 36 distinct fields, listed in Table\,\ref{targets}, was surveyed with the VLA at 6, 15,
and 22\,GHz in the A- and B-array configurations. In \citet{Moscadelli2016}, we presented a pilot study of 11 targets belonging to the
POETS survey (hereafter, Paper\,I). More details about the target selection can be found in Paper\,I. Here, we present the full sample of radio
continuum sources; a following paper will combine continuum and maser information.

In Sections~2 and~3, we summarize the observation information and explain the details of the data analysis, respectively. In Section~4,
we present and discuss the overall results of the radio continuum survey. Conclusions are drawn in section~5.

%\clearpage

%_____________________________________________________________
%                                                    TABLES  # 1
%-----------------------------------------------------------------------------------------------------------

%\addtocounter{table}{0}
%\begin{landscape}
\begin{table}
\caption{Target fields of the POETS survey.}\label{targets}
\centering
\begin{tabular}{l r r r}

\hline \hline
         &     \multicolumn{2}{c}{Phase Center}        &  \\
\multicolumn{1}{c}{Field} &  \multicolumn{1}{c}{R.A.\,(J2000)}  &       \multicolumn{1}{c}{Dec.\,(J2000)}        &          \multicolumn{1}{c}{L$_{\rm H_2O}$}         \\  
         &    \multicolumn{1}{c}{(h\,m\,s)}      &  \multicolumn{1}{c}{($^{\circ}$\,$'$\,$''$)}  &    \multicolumn{1}{c}{(10$^{-5}$\,L$_{\odot}$)}  \\  
\hline
G009.99$-$0.03  & 18:07:50.100 & $-$20:18:56.00 & 0.75 \\
G012.43$-$1.12  & 18:16:52.100	& $-$18:41:43.00 & 4.98  \\
G012.90$-$0.24  & 18:14:34.400 & $-$17:51:52.00 & 0.36  \\
G012.91$-$0.26  & 18:14:39.400 & $-$17:52:06.00 & 0.05 \\
G014.64$-$0.58  & 18:19:15.500 & $-$16:29:45.00 & 0.10 \\
%G024.49$-$0.04  & 18:36:05.700 & $-$07:31:19.00 &  ... \\
G026.42$+$1.69 & 18:33:30.500 & $-$05:01:02.00 & 0.31 \\
G031.58$+$0.08 & 18:48:41.700 & $-$01:09:59.00 & 0.77 \\
G035.02$+$0.35 & 18:54:00.700 & $+$02:01:19.00 & 0.63 \\
%G045.07$+$0.13 & 19:13:22.000 & $+$10:50:53.00 & ... \\
%\textbf{G048.61$+$0.02} & 19:20:31.200 & $+$13:55:25.00 & ... \\
G049.19$-$0.34  & 19:22:57.800 & $+$14:16:10.00 & 9.87 \\
G076.38$-$0.62  & 20:27:25.500 & $+$37:22:48.00 & 0.24 \\ 
G079.88$+$1.18 & 20:30:29.100 & $+$41:15:54.00 & 0.08 \\  
G090.21$+$2.32 & 21:02:22.700 & $+$50:03:08.00 & 0.03 \\  
G105.42$+$9.88 & 21:43:06.500 & $+$66:06:55.00 & 0.16 \\   
G108.20$+$0.59 & 22:49:31.500 & $+$59:55:42.00 & 3.61 \\
G108.59$+$0.49 & 22:52:38.300 & $+$60:00:52.00 & 0.15 \\
G111.24$-$1.24  & 23:17:20.800 & $+$59:28:47.00 & 1.06 \\
G160.14$+$3.16 & 05:01:40.200 & $+$47:07:19.00 & 0.41 \\
G168.06$+$0.82 & 05:17:13.700 & $+$39:22:20.00 & 14.32 \\
G176.52$+$0.20 & 05:37:52.100 & $+$32:00:04.00 & 0.14 \\
G182.68$-$3.27  & 05:39:28.400 & $+$24:56:32.00 & 0.41 \\            
G183.72$-$3.66  & 05:40:24.200 & $+$23:50:55.00 & 1.51 \\
G229.57$+$0.15 & 07:23:01.800 & $-$14:41:36.00 & 0.71 \\
G236.82$+$1.98 & 07:44:28.200 & $-$20:08:30.00 & 0.51 \\  
G240.32$+$0.07 & 07:44:51.900 & $-$24:07:43.00 & 1.24 \\  
G359.97$-$0.46  & 17:47:20.200 & $-$29:11:59.00 & 1.57 \\
%  & & & ... \\
 & & &  \\
 \multicolumn{4}{c}{Targets presented in \citet{Moscadelli2016}}  \\
 & & &  \\
 G005.88$-$0.39  & 18:00:30.306 & $-$24:04:04.48 & 1.86 \\
 G011.92$-$0.61  & 18:13:58.120 & $-$18:54:20.28 & 3.27 \\
 G012.68$-$0.18  & 18:13:54.744 & $-$18:01:47.57 & 8.83 \\
 G016.58$-$0.05  & 18:21:09.084 & $-$14:31:49.56 & \tablefootmark{a}4.36 \\
 G074.04$-$1.71  & 20:25:07.104 &    $+$34:49:58.58 & 0.25 \\
 G075.76+0.34     & 20:21:41.086 &    $+$37:25:29.28 & 0.75 \\
 G075.78+0.34     & 20:21:41.086 &    $+$37:25:29.28 & 7.98 \\
 G092.69+3.08     & 21:09:21.724 &    $+$52:22:37.10 & 2.98 \\
 G097.53+3.18     & 21:32:12.441 &    $+$55:53:50.61 & 69.72 \\
 G100.38$-$3.58  & 22:16:10.368 &    $+$52:21:34.11 & 1.40 \\
 G111.25$-$0.77  & 23:16:10.360 &    $+$59:55:29.53 & 1.36 \\
\hline
\end{tabular}

% In this Table, include also a note for those two sources which come from program 12B-044.

\tablefoot{Column\,1: field name. Columns\,2 and~3: phase center of the observations. Column\,4: isotropic H$_2$O maser luminosity.
For each source, this luminosity refers to the averaged luminosity measured with the VLBA in the period from March 2011 to April 2012
(from the BeSSeL Survey data). The isotropic maser  luminosity was estimated with the following formula: 
(L$_{\rm H_2O}/$L$_{\odot}$)\,$=$\,2.3\,$\cdot$\,10$^{-8}$\,$\times$\,(S$_{\rm H_2O}$\,Jy\,km\,s$^{-1}$)\,$\times$\,(D$/$kpc)$^2$,
where S$_{\rm H_2O}$ and D are, respectively, the integrated H$_2$O maser flux and the heliocentric distance to the H$_2$O maser site.
\tablefoottext{a}{Luminosity estimated from data in \cite{Sanna2010a}.}}
\end{table}
%\end{landscape}

%_____________________________________________________________
%-----------------------------------------------------------------------------------------------------------

\section{Observations and calibration}\label{obs}

We conducted VLA\footnote{The National Radio Astronomy Observatory is a facility of the National Science Foundation operated under
cooperative agreement by Associated Universities, Inc.} observations, under program 14A-133, at C, Ku, and K bands towards 26 distinct
Galactic fields associated with strong H$_2$O maser emission (Table\,\ref{targets}). Towards each field, we observed with the A-array
configuration at C and Ku bands, and with the B-array configuration at K band, in the periods March--May 2014 and February-March 2015,
respectively.  

%_____________________________________________________________
%                                                    TABLES  # 2
%-----------------------------------------------------------------------------------------------------------

%\addtocounter{table}{0}
%\begin{landscape}
\begin{table*}
\caption{Summary of VLA observations (code 14A-133).}\label{settings}
\centering
\begin{tabular}{c c c c c c c c c }

\hline \hline

 Band   &  $\nu_{\rm center}$   &  BW     &  Array  & On-source  & $\theta_{\rm LAS}$ & HPBW  &            RMS noise               & RMS T$_b$         \\  
           &             (GHz)             &  (GHz)  &           &   (min)       &     ($''$)                   & ($''$)  &  ($\mu$Jy\,beam$^{-1}$)  &        (K)               \\  

\hline
 C   & 6.0   & 4.0 & A  & 15  & 8.9 & 0.33 &  8.0 &  3.0 \\
 Ku & 15.0 & 6.0 & A  & 10  & 3.6 & 0.13 &  9.0 &  3.0 \\
 K   & 22.2 & 8.0 & B  & 15  & 7.9 & 0.29 & 11.0 & 0.3  \\

%& & & & & & & & &  \\
\hline
%& & & & & & & & &  \\
\end{tabular}

\tablefoot{Columns\,1, 2, and 3: radio band, central frequency of the observations, and receiver bandwidth used. Columns\,4 and 5: array configuration
and  on-source integration time at each band. Column\,6: maximum recoverable scale of the emission at a given frequency. \emph{The following values
were derived from the VLA Exposure Calculator (v.18A)}. Column\,7: approximate synthesized beam size obtained with Robust weighting. Columns\,8
and 9: expected thermal rms noise and corresponding  brightness temperature. The rms values were estimated assuming a 15\,\% loss of bandwidth at
each frequency.}
\end{table*}
%\end{landscape}

%_____________________________________________________________
%-----------------------------------------------------------------------------------------------------------

For high-sensitivity continuum observations in the C, Ku, and K bands, we employed the 3-bit samplers observing dual polarization over  
the largest receiver bandwidth of 4, 6, and 8 GHz, respectively. At each band, we made use of the wideband setup, tuning, respectively,
2, 3, and 4 2-GHz wide IF pairs across the receiver bandwidth. Each IF comprised 16, 128-MHz wide subbands with a channel spacing 
of 1\,MHz. At C and K bands, we also used a narrow spectral unit of 4\,MHz centered on the 6.7\,GHz methanol and 22\,GHz water
maser lines. These bands had widths of 180 and 54\,km\,s$^{-1}$ to cover the respective maser emission. In addition, they were
correlated with 1664 and 128 channels to achieve sufficient velocity resolutions of 0.1 and 0.4\,km\,s$^{-1}$ to spectrally resolve
single maser lines. In the following, we report on the radio continuum data.
 
Our previous pilot observations, conducted under program 12B-044 \citep{Moscadelli2016}, have demonstrated that the free-free continuum
emission, associated with the H$_2$O masers, has peak flux densities of a few 100\,$\mu$Jy at centimeter wavelengths, on average. Therefore,
we integrated on each source for 10-20 minutes, in order to achieve a thermal rms noise of the order of 10\,$\mu$Jy\,beam$^{-1}$,
and a signal-to-noise ratio in excess of 10, typically. More details on the observational strategy can be found in Section\,3.1 of 
\citet{Moscadelli2016}. Observation information is summarized in Table\,\ref{settings}.

Data reduction was performed within the Common Astronomy Software Applications package (CASA), making use of the VLA pipeline. Radio 
fluxes were calibrated with the Perley-Butler 2013 flux density scale. Additionally, we flagged each dataset based on the quality of both the
amplitude and phase calibration, and performed self-calibration on those radio continuum maps which were limited in dynamic range 
(10\,$<$\,SNR\,$<$\,50), corresponding to source fluxes greater than a few 100\,$\mu$Jy.

\section{Method}\label{analysis}

For each field, in Figs.\,\ref{fig1}--\ref{fig5} we present the radio continuum maps in the C, Ku, and K bands and analyse the radio
spectral index of the continuum emission following two methods: by integrating the radio fluxes at each band and fitting the spectral slope
among the bands (Sect.\,\ref{sed}); by comparing the radio emission among the bands in the \textit{uv}-plane directly (Sect.\,\ref{amap}),
making use of the multi-frequency-synthesis (MFS) deconvolution of the task \emph{clean} of CASA \citep{Rau2011}.  In the following, we
provide the details of our analysis. 

\subsection{Continuum maps}\label{cmap}

At variance with our pilot program 12B-044 (Paper\,{\sc I}), we observed at K band with the B-array configuration instead of the A-array,
so that the C- and K-band observations are sensitive to the same angular scales, covering a common range of spatial frequencies
($<$\,900\,k$\lambda$). The observations at Ku band cover greater \textit{uv}-distances in excess of 2000\,k$\lambda$. Continuum
maps are shown in the left panels of Figs.\,\ref{fig1}--\ref{fig5} for each frequency band. Imaging remarks are listed in the following: 

%\begin{description}[\noindent]
\vspace{0.10cm}

\noindent \textbf{Beam \& weighting.} Each field was imaged with the task \emph{clean} of CASA interactively and with a circular restoring beam
(column\,4 of Table\,\ref{contab}). The restoring beam size was set equal to the geometrical average of the major and minor axes of the clean
beam size. The pixel size is a factor 0.2--0.25 of the half-power-beam-width (HPBW). By rule of thumb, the C- and K-band maps were produced
with a common Briggs robustness parameter of 0.5, to simultaneously optimize side-lobes suppression and sensitivity; the Ku-band maps were
produced with natural weightings to enhance the sensitivity in the shortest baselines. For faint sources with peak brightness below 7\,$\sigma$,
we applied natural weightings without distinction. The beam size ranges from a minimum of $0\farcs110$, at Ku band, to a maximum $0\farcs481$,
at C band.

\noindent \textbf{UV-range.} At the bottom of each plot,  we indicate with ``UVCUT'' or ``TAPER'' whether a lower limit to the \textit{uv}-range
or tapering were used for cleaning, respectively. At each frequency, the targets in our sample are more compact than the largest angular size
($\theta_{\rm LAS}$) which can be imaged with the A- and B-array configurations (Table\,\ref{settings}). On the other hand, 5 fields in our
sample exhibit extended bright continuum emission (other than the targets) which cannot be recovered without shorter baselines. Since this
emission adds to the image residuals, and limits the image sensitivity, we cleaned those fields by setting a lower cut to the \textit{uv}-plane. A
lower cut of 100\,k$\lambda$ corresponds to filter out emission more extended than $2''$. Details are given in the footnotes to Table\,\ref{contab}.

\noindent \textbf{Field-of-view.} In Figs.\,\ref{fig1}--\ref{fig5}, we limited the field-of-view so as to include the radio continuum sources closest to
the H$_2$O maser emission detected in the field. For the brightest maser spots ($\ga$\,0.5\,Jy\,beam$^{-1}$), absolute positions can be
obtained from the Bar and Spiral Structure Legacy (BeSSeL) website at \url{http://bessel.vlbi-astrometry.org/first_epoch} \citep{Reid2014}. In
each plot, we draw a scale bar to quantify the linear size of the field; the linear size was computed from the distance values reported in
Table\,1 of Paper\,{\sc I}. Field-of-views range from a minimum of 650\,au (Fig.\,\ref{fig2}) to a maximum of 9000\,au  (Fig.\,\ref{fig1}).

\vspace{0.10cm}
%\end{description}

For each field, we superpose, where available, the C-band emission in black contours (and gray tones), the Ku-band emission in red contours,
and the K-band emission in blue contours. Map contours are listed at the bottom of each plot and are scaled at multiples of the 1\,$\sigma$
rms (column\,5 of Table\,\ref{contab}). The rms noise was estimated with the task \emph{imstat} of CASA, across a region without continuum emission.
Alternatively, we indicate whether the field was not observed at a given frequency band, or the emission was not detected above a 3--5\,$\sigma$ level. 

When multiple radio continuum peaks exceeding 5\,$\sigma$ are detected (e.g., Fig.\,\ref{fig1}), individual peaks are labeled with numbers increasing
with decreasing brightness. For each peak or ``component'' in Table\,\ref{contab}, we report the maximum pixel position (columns\,6--7), the
maximum pixel value (column\,8), and the integrated flux (column\,10) obtained with the task \emph{imstat} of CASA. Values of integrated flux
are measured in images produced with a common \textit{uv}-coverage among the bands (see Sect.\,\ref{sed}). The flux density of single components was computed
within the 3\,$\sigma$ contour at each band. Blended sources were detected in 2 fields (Figs.\,\ref{fig1} and~\ref{fig3}), and their flux densities
were computed within squared boxes centered on each peak.

In column\,9 of Table\,\ref{contab}, we assign a grade to each source depending on the compactness of the emission. We fitted the continuum 
emission within the 3\,$\sigma$ contour with a two-dimensional (elliptical) Gaussian (task \emph{imfit} of CASA), and classified the source 
size as follows: 
the source is compact (``C'') if its deconvolved size is smaller than half the beam size; 
the source is slightly resolved (`SR'') if its deconvolved size is smaller than the beam and larger than half the beam size; 
the source is resolved (``R'') if its deconvolved size is larger than the beam size.

%Tapering at 500\,k$\lambda$ was used at Ku-band for two fields, G009.99$-$0.03 G079.88$+$1.18. ---> inlude this info in the table directly. 

\subsection{Spectral energy distributions}\label{sed}

In Figs.\,\ref{fig1}--\ref{fig5}, we show the results of the spectral index analysis based on the integrated fluxes at each band (SED), under the 
assumption that: $S_{\nu_1}/S_{\nu_0}$\,$=$\,$(\nu_1/\nu_0)^{\alpha}$. As a caveat to the spectral index analysis, we assume there was
no significant source variability during the one-year interval between the A- and B-array configurations. Similar to Paper\,{\sc I}, we measured the
continuum spectral index following four steps:

\begin{enumerate}

\item First, we constrained a common \textit{uv}-range among the radio bands of 40--800\,k$\lambda$, and imaged each band separately
with the same round beam ($0\farcs30$) and pixel size ($0\farcs06$). Higher values of the minimum \textit{uv}-distance were used for 
4 fields (see Sect.\,\ref{cmap}). On top of each plot, we specify the range of \textit{uv}-distances selected for a given field.

\item These maps were used to select a polygon around the source which encloses the continuum emission; these polygons typically
correspond to the 3\,$\sigma$ contours. For instance, sources detected at C, Ku, and K bands define three distinct polygons. At a given frequency, 
we estimated the integrated flux as the averaged flux over the three polygons. The uncertainty of the averaged flux was estimated as the dispersion
of the mean, and was summed in quadrature with a $10\%$ error in the absolute flux scale. This method takes into account that, due to small
position shifts among the C-, Ku-, and K-band maps ($\leq0\farcs1$), the same polygon does not trace the same region at different frequencies.
% Integrated fluxes of blended sources were computed within a same squared box at each frequency. 
% Lower limits are the same defined for the continuum maps with full \textit{uv}-range.

\item We used the averaged fluxes and their uncertainties to compute the continuum spectral index with a linear regression fit,
where: $\log_{10}(S_{\nu})=\alpha\,{\cdot}\,\log_{10}(\nu)+const.$ For sources only detected at K band, we provide a lower
limit to the spectral index, by fitting a straight line to the K-band flux and the upper limit of the Ku-band flux.

\item  We plot the averaged fluxes at the C, Ku, and K bands (Fig.\,\ref{fig1}--\ref{fig5}); each frequency is identified with the same color used for the
continuum maps.  In each plot, we draw the spectral index values (and uncertainties) of the different band combinations, as well as the spectral index
among the three bands where available. 

\end{enumerate}

In column\,11 of Table\,\ref{contab}, the best-fitting value of the spectral index is given for each source. 

%We observe a systematic loss of flux at Ku band in ...\% of the sample, with respect to the flux expected by interpolating among 
%the C, Ku, and K bands together (e.g., field \# ... in Table\,\ref{contab} and Fig.\,\ref{fig4}). We interpret this difference as due
%to the higher number of long baselines, with an equal \textit{uv}-coverage, probed by the Ku-band observations with respect to
%the C and K bands. For two fields of the sample, which were only observed  at Ku and K bands, we divided the K band in two
%sub-bands, each one equal to half the bandwidth, in order to derive the spectral index  between 18 and 26\,GHz, and compare with
%that between the Ku and K bands (fields \# ... in Table\,\ref{contab}). 

\subsection{Spectral index maps}\label{amap}

\begin{figure*}
%\sidecaption
%\centering
\includegraphics [angle= 0, scale= 1.0]{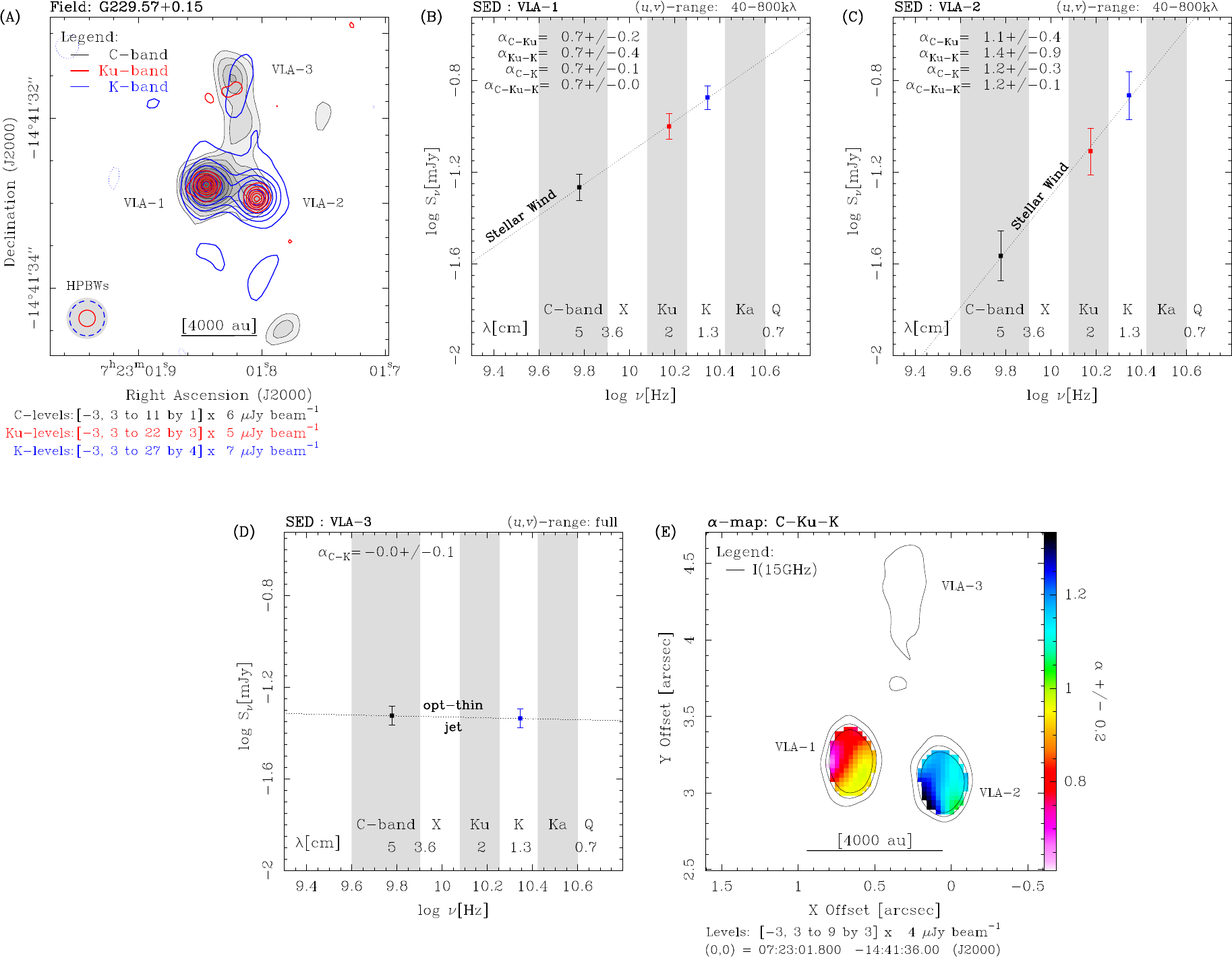}
\caption{Example of stellar wind emission from a double system: H$_{2}$O maser site G229.57$+$0.15 (Sect.\,\ref{ex1}). 
\textbf{Panel\,A:} superposition of the VLA maps for the C- (grey-scale/black contours), Ku- (red), and K-band (blue) emission
(Sect.\,\ref{cmap}).  Contour levels, at multiples of the 1\,$\sigma$ rms, are indicated in the footnote. Synthesized beams for each
band are shown in the bottom left corner. A scale bar in units of au is drawn near the bottom axis. Three distinct continuum sources are
identified and labeled VLA--1, VLA--2, and VLA--3, from the brightest to the faintest.
\textbf{Panels\,B--D:} spectral energy distribution for the three sources identified in panel\,A separately (Sect.\,\ref{sed}). Each plot
shows the logarithm of  the integrated flux (in mJy) as a function of the logarithm of the observed frequency (in Hz). Different
frequency bands are labeled near the bottom axis together with the reference wavelength; grey shades mark the boundary of each band.
For each band combination, the linear spectral index value ($\alpha$) and its uncertainty are specified in the upper left corner. The common 
range of of \textit{uv}-distances (e.g., 40--800\,k$\lambda$ in panel B, and full \textit{uv}-range in panel D), used to compute the
integrated fluxes among the bands, is indicated in the upper right.  
\textbf{Panel\,E:} color map of the linear spectral index ($\alpha$) computed with the multi-frequency-synthesis algorithm of the
task \emph{clean} of CASA (Sect.\,\ref{amap}). Frequency bands used to fit the spectral index are listed on top; spectral index values
are quantified by the right-hand wedge. The uncertainty on $\alpha$ is indicated in the wedge label; this value is an upper limit which holds at 
the 7\,$\sigma$ contour of the average brightness map (black contours). Contour levels of the average brightness map, at multiples
of the 1\,$\sigma$ rms, are indicated in the footnote; the central frequency ($\nu_0$) of the average brightness map is specified in
the legend. The absolute coordinates of (0,0) offset are indicated in the footnote (see Table\,\ref{targets}). Radio continuum components and
the scale bar as in panel\,A.}\label{fig1}
\end{figure*}

\begin{figure}
%\sidecaption
\centering
\includegraphics [angle= 0, scale= 1.0]{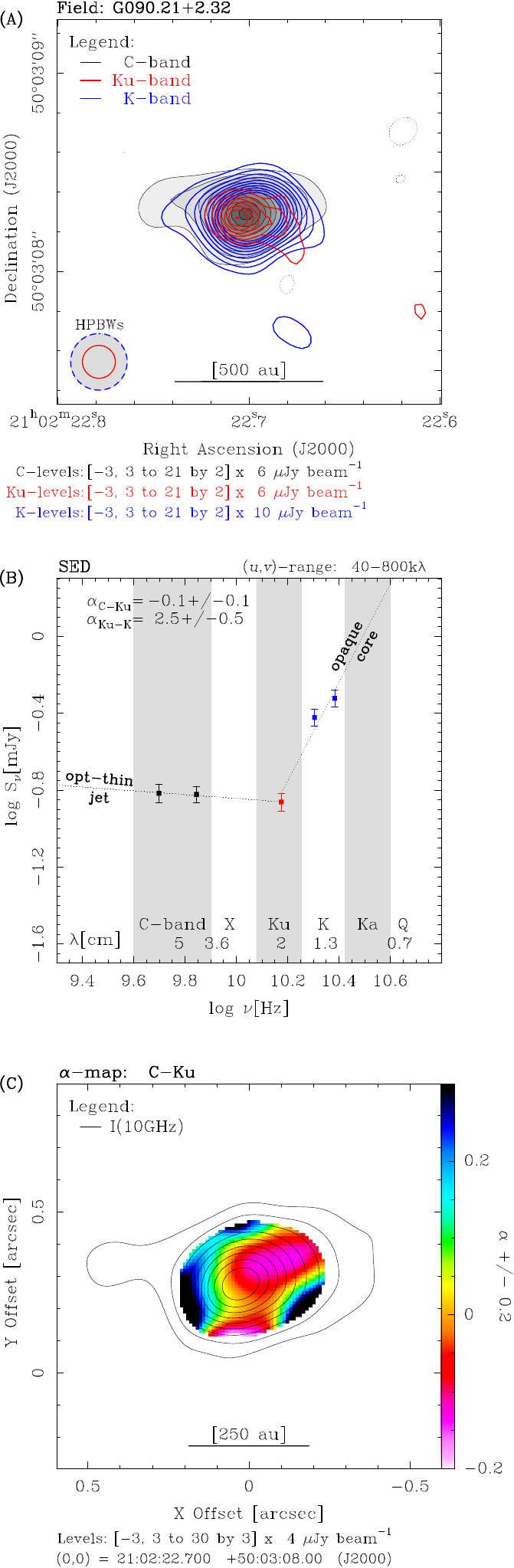}
\caption{Example of optically-thin thermal jet emission: H$_{2}$O maser site G090.21$+$2.32 (Sect.\,\ref{ex2}). Labels and symbols as in Fig.\,\ref{fig1}. 
In panel\,B, we have divided the C and K bands in two sub-bands, in order to show that the spectral index within each band is consistent with the spectral 
index interpolated to the Ku-band frequency. The spectral energy distribution shows a region of optically thin (resolved) emission below 15\,GHz, and 
a region of optically thick (compact) emission above 15\,GHz. The spectral index map of panel\,C was computed within the (linear) range of 
optically thin emission (C and Ku bands only).}\label{fig2}
\end{figure}

In Figs.\,\ref{fig1}--\ref{fig5}, we provide an alternative analysis of the continuum spectral index based on the MFS cleaning
(\citealt{Rau2011}, and references therein), under the assumption that: $I_{\nu_1}/I_{\nu_0}$\,$=$\,$(\nu_1/\nu_0)^{\alpha}$.
The task \emph{clean} of CASA was run with parameter \emph{nterm\,$=$\,2}, which expands the observed brightness distribution
taking into account the first two terms of a Taylor series: $I_{\nu}=I_{\nu_0}+{\alpha}\,{\cdot}\,I_{\nu_0}\,{\cdot}\,(\nu-\nu_0)$. This procedure 
produces a set of Taylor-coefficient images (Sects.\,2.2 and~2.7 of \citealt{Rau2011}): the first coefficient of the Taylor expansion
defines the average brightness distribution (in CASA, map extension ``$.tt0$''); the spectral index map is derived from the second
coefficient  (in CASA, map extension ``$.alpha$''), which is the product of the average brightness times the local spectral index value. 
The MFS algorithm compares observed and modeled visibilities in the \textit{uv}-plane, providing information on the spatial distribution
of the spectral index value, as opposed to the average spectral index inferred in Sect.\,\ref{sed}. Hereafter, we refer
to the this method as the ``$\alpha$-map''.

Since the accuracy of the spectral index calculation depends on the SNR of the brightness distribution, the following constraints apply:

\begin{enumerate}

\item $\alpha$-maps were only processed for those sources where the continuum emission exceeds 7\,$\sigma$ at each band;

\item $\alpha$-maps are only displayed where these two conditions hold: the average map exceeds 7\,$\sigma$, and
the uncertainty of $\alpha$ estimated by the algorithm is less than 0.2--0.3 (map extension ``$.alpha.error$'').

\end{enumerate}

For each source which satisfies the first requirement, we cleaned the datasets available all together with a Briggs robustness parameter
of 0.5. In Figs.\,\ref{fig1}--\ref{fig5}, each $\alpha$-map shows the average brightness map in contours and the spectral index map
in colors, according to the right-hand wedge. In each plot, we also specify the center frequency of the average brightness map, and the
uncertainty on $\alpha$, which increases from the brightness peak, where $\Delta\alpha<$0.1 typically, to the outer iso-contour,  
where $\Delta\alpha=$\,0.2 or 0.3. 

In Appendix\,\ref{spvar}, we show that small position shifts, between maps at different frequency bands, produce systematic 
variations in the spectral index maps. Since these position shifts are of the same order of the calibration uncertainties, one has to 
exert caution when interpreting spectral index gradients.

%_____________________________________________________________
%                                              FIGURES  N.1
%-----------------------------------------------------------------------------------------------------------

%_____________________________________________________________
%-----------------------------------------------------------------------------------------------------------
%_____________________________________________________________
%                                              FIGURES  N.2
%-----------------------------------------------------------------------------------------------------------

%_____________________________________________________________
%-----------------------------------------------------------------------------------------------------------

\section{Results and discussion}\label{results}

\begin{figure*}
%\sidecaption
%\centering
\includegraphics [angle= 0, scale= 0.95]{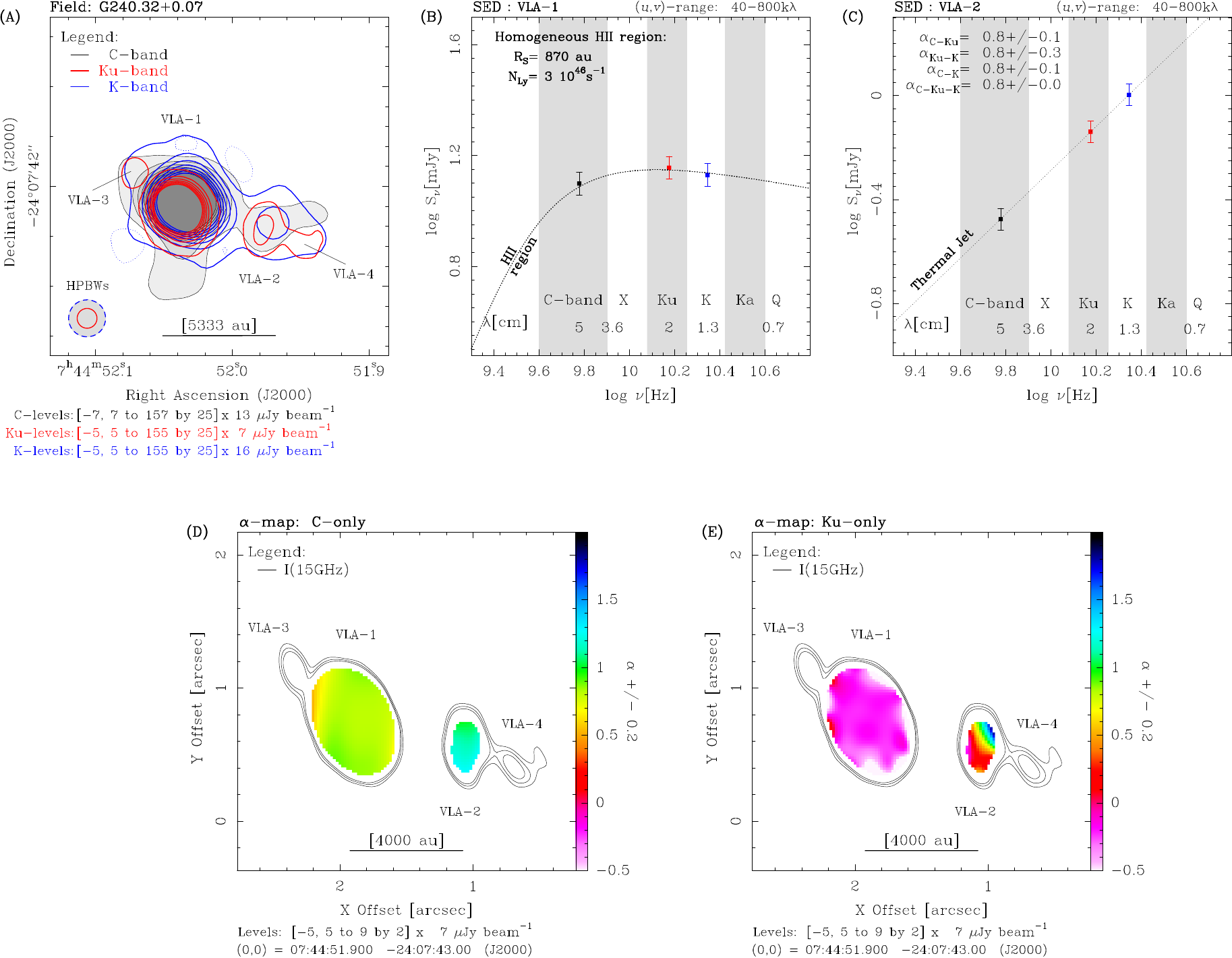}
\caption{Example of radio thermal jet (VLA--2) near to a HCH\,{\sc ii} region (VLA--1): H$_{2}$O maser site G240.32$+$0.07  (Sect.\,\ref{ex3}). 
Labels and symbols as in Fig.\,\ref{fig1}. The spectral energy distribution of source VLA--1 was fitted with a model of homogeneous H\,{\sc ii} region (panel\,B),
with Str\"{o}mgren radius of 870\,au and number of Lyman photons of 3\,$\times$\,10$^{46}$\,s$^{-1}$. This model accurately reproduces the observed
fluxes at each band. Thanks to the high signal-to-noise ratio, in panels~D and~E we can determine the spectral index maps of the C and Ku bands, separately,
with an uncertainty better than $\pm$\,0.2. The average brightness map (contours) is that of the Ku band in both panels~D and~E, for comparison. The spectral
index maps trace the turn-over of the continuum spectrum for source VLA--1, from $\alpha$\,$=$\,1.0--6.0,  between 4 and 8\,GHz (panel\,D), to
$\alpha$\,$=$\,--0.1, above 15\,GHz (panel\,E).}\label{fig3}
\end{figure*}

In this section, we firstly discuss the details of four examples of radio continuum emission  (Figs.\,\ref{fig1}--\ref{fig4}), which
provide the background to interpret the entire sample of sources (Fig.\,\ref{fig5}). We then comment on the nature of the radio 
continuum emission by comparing the radio luminosity of our sample with both the H$_2$O maser and bolometric luminosities.

In the following, if the radio continuum SED is well reproduced with a linear spectral slope, we will use the terms (ionized)
\emph{stellar wind} and \emph{thermal jet} to distinguish between unresolved and resolved/elongated emission,
respectively. In column\,12 of Table\,\ref{contab}, we provide a classification for the entire sample of sources, which split as 
follows: 15 stellar winds, 8 thermal jets, 2 H\,{\sc ii} regions, and 7 sources detected at one frequency only (and not classified).

\subsection{Example\,1: compact stellar winds from a double system}\label{ex1}

In Fig.\,\ref{fig1}, we present an example of stellar wind sources towards the H$_2$O maser site G229.57$+$0.15. 
In panel\,A, we plot the brightness distribution of the continuum emission detected in the three observed bands. These maps
show two compact sources, labeled VLA--1 and VLA--2, which are aligned along the east-west direction, and a region of extended 
(resolved) emission, labeled VLA--3, which extends northwards of VLA--1 and VLA--2. The radio continuum peaks of
VLA--1 and VLA--2 are separated by about 2800\,au (Table\,\ref{contab}). These maps were produced with natural weightings,
to retrieve the extended flux associated with VLA--3, and have approximately the same rms noise of 5--7\,$\mu$Jy\,beam$^{-1}$.

Radio continuum components VLA--1 and VLA--2 have comparable brightness at K band of 170\,$\mu$Jy\,beam$^{-1}$, but the
peak  of VLA--2  is about half that of VLA--1 at C-band (Table\,\ref{contab}). This difference determines 
a much steeper gradient in the SED of VLA--2 with respect to that of  VLA--1, which is evident by comparing panels B and C of
Fig.\,\ref{fig1}. In both panels, the integrated fluxes computed at each frequency are strictly aligned, and define a linear
gradient ($\alpha$) between the C, Ku, and K bands of 0.68\,$\pm$\,0.02 and 1.2\,$\pm$\,0.1, for the VLA--1 and VLA--2 sources
respectively. The higher flux uncertainties of VLA--2, with respect to VLA--1, are due to the larger difference between the fluxes
integrated within the C-, Ku-, and K-band boxes. These boxes coincide with the 3\,$\sigma$ contours everywhere except between VLA--1
and VLA--2, where we drew a clear cut along the north-south direction to separate the emission. The spectral index values of VLA--1
and VLA--2 are consistent with the radio continuum emission being produced by thermal bremsstrahlung from a stellar wind.

At variance with sources VLA--1 and VLA--2, source VLA--3 is resolved out at the Ku band, and, to derive the continuum spectral index
in this region, we made use of the C and K bands only (panel\,D). Since the C and K bands observations are sensitive to the same
spatial scales, we did not limit the \textit{uv}-range to compute the integrated fluxes, and used a squared box which encloses the 
emission between a declination of --14:41:32.03 and --14:41:31.17. The spectral index value of VLA--3 is of 0.0\,$\pm$\,0.1, and,
taking into account the spatial morphology of the radio continuum emission, it is consistent with thermal bremsstrahlung from an
optically thin radio jet. This emission might be associated either with VLA--1 or VLA--2.

In panel\,E, we plot the spectral index map determined from the combination of the C-, Ku-, and K-band datasets; this map is \textbf{referred} 
to the central frequency of the Ku band (15\,GHz). Only sources VLA--1 and VLA--2 have sufficiently high SNR ($>$\,7\,$\sigma$) to
compute the spectral index in the \textit{uv}-domain with high accuracy. The color scale of panel\,E shows that the spectral index
of VLA--1 ranges from 0.65 to 0.95, encompassing the average value determined in panel\,B. The extremes of the alpha range 
coincide with the 7\,$\sigma$ contour of the average brightness map (black contours), where the uncertainty on $\alpha$ increases to
$\pm$\,0.2. Similarly, the spectral index map of VLA--2 converges, towards the brightness peak, on the average value determined in
panel\,C; the spectral index varies over an interval 1.0--1.3 of about 1\,$\sigma$.

Notably, sources VLA--1 and VLA--2 make the case for two nearby (unresolved) stellar winds with sharply different spectral indexes.
Following \citet{Reynolds1986}, the higher spectral index value of VLA--2 may be interpreted as the radio continuum opacity changing
more rapidly in VLA--2 than in VLA--1. We make use of eq.\,15 of \citet{Reynolds1986}, which quantifies the spectral index of a radio
jet as a function of its geometry ($\epsilon$), temperature (q$_T$), and optical depth ($q_{\tau}$) variations across the flow. We 
assume that the two flows are either spherical or conical ($\epsilon$\,$=$\,1), and since both sources are unresolved at the beam scale,
we take an upper limit to the source size of $0\farcs206$ at K band, corresponding to 930\,au at a distance of 4.52\,kpc. We also
assume that the ionized gas is approximately isothermal within a radius of 500\,au from the central source (q$_T$\,$=$\,0). Under
these conditions, eq.\,15 of \citet{Reynolds1986} predicts that:   
($\alpha_1$--$\alpha_2$)$/4.2$\,$=$\,$\rm (1/q_{\tau_1})$\,--\,$\rm (1/q_{\tau_2})$\,$=$\,--\,0.12, where q$_{\tau_1}$ and
q$_{\tau_2}$ are negative parameters, and the subscripts~1 and~2 are relative to sources VLA--1 and VLA--2, respectively.
This calculation implies that the absolute value of q$_{\tau_1}$ is smaller than that of q$_{\tau_2}$.

%_____________________________________________________________
%                                              FIGURES  N.3
%-----------------------------------------------------------------------------------------------------------

%_____________________________________________________________
%-----------------------------------------------------------------------------------------------------------

\subsection{Example\,2: optically-thin thermal jet emission and opaque core}\label{ex2}

In Fig.\,\ref{fig2}, we present an example of radio thermal jet source towards the H$_2$O maser site G090.21$+$2.32. This source 
has the lowest bolometric luminosity in our sample (27\,L$_{\odot}$). The radio continuum maps show that the
emission is elongated in the east-west direction (panel\,A). At the C and Ku bands (4--18\,GHz), the continuum emission is resolved
at a scale of $0\farcs292$ and $0\farcs166$, respectively; at the K band (18--26\,GHz), the continuum  emission has a deconvolved 
size of the order of half the beam ($0\farcs140$). At a distance of 670\,pc, the source size corresponds to about 90\,au in the higher
frequency range, and exceeds 200\,au in the lower frequency range.

The flux density at C band is comparable to that at the Ku band, which is almost three times less than the flux recovered at K band
(Table\,\ref{contab}). To verify that the difference between the Ku- and K-band fluxes does not depend on the calibration, we divided 
both the C and K bands in two sub-bands, each one equal to half the bandwidth, and imaged each sub-band separately. The integrated 
fluxes derived from each sub-band are plotted in panel\,B of Fig.\,\ref{fig2}. In this plot, we show that the spectral index fitted 
between the C and Ku bands reproduces the spectral slope determined within the C band only; similarly, the spectral slope within the 
K band only is consistent with that between the Ku and K bands. The spectral index values derived between the C and Ku bands 
($-0.1$\,$\pm$\,0.1), and between the Ku and K bands (2.5\,$\pm$\,0.5), are consistent, respectively, with thermal jet emission 
which is optically thin for frequencies below 15\,GHz, and with a compact core which is optically thick for frequencies above 15\,GHz.
This scenario closely resembles that predicted by \citet[e.g., their Fig.\,2]{Reynolds1986}.

In panel\,C, we plot the spectral index map determined from the combination of the C- and Ku-band datasets; this map is referred 
to the central frequency between the C and Ku bands (10\,GHz). The spectral index map ranges from $-0.2$ to 0.3; at these extremes,
corresponding to the outer iso-contours, the uncertainty increases to $\pm$\,0.2, and decreases below $\pm$\,0.1 at the brightness
peak. This range of spectral index values is consistent with the average spectral index derived in panel\,B, within 1\,$\sigma$. Spectral
index variations observed in panel\,C are likely an effect of the MFS deconvolution algorithm, produced by combining different Gaussian
distributions at distinct frequency bands (see Appendix\,\ref{spvar}).

%_____________________________________________________________
%                                              FIGURES  N.4
%-----------------------------------------------------------------------------------------------------------
\begin{figure*}
%\sidecaption
%\centering
\includegraphics [angle= 0, scale= 0.95]{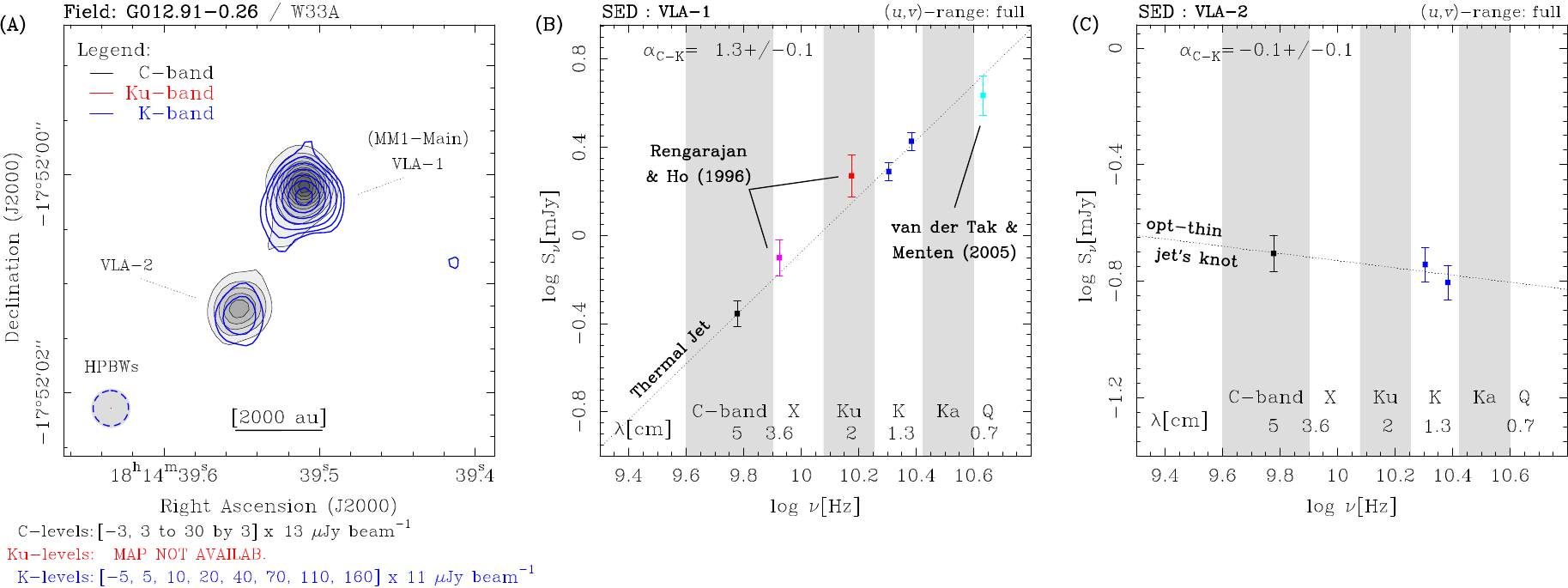}
\caption{Example of radio thermal jet (VLA--1) with nearby knot (VLA--2): H$_{2}$O maser site G012.91$-$0.26 or W33A (Sect.\,\ref{ex4}). 
Labels and symbols as in Fig.\,\ref{fig1}. Two radio continuum sources are detected at both C and K bands along a position angle of approximately 
$152\degr$ (panel\,A): the brightest emission coincides in position with the brightest millimeter source in the dusty core MM1. In panels\,B and~C,
we have fitted the spectral index of sources VLA--1 and VLA--2 with the measured fluxes at C and K bands, after dividing the K-band observations
in two sub-bands, each spanning 4\,GHz. In panel\,B, fluxes reported in the literature at X (pink), Ku (red), and Q (cyan) bands are consistent with a
single linear slope, from 4 to 43\,GHz. The radio continuum emission associated with W33A--MM1 can be interpreted as a thermal jet, whose
components VLA--1 and VLA--2 mark, respectively, the ionized gas closest to the driving source, namely MM1-main, and a knot of shocked gas along
the jet direction.}\label{fig4}
\end{figure*}

%The spectral index of source VLA--1 is in the range expected for thermal bremsstrahlung from a stellar wind or jet. In panel\,C, the spectral index
%of source VLA--2, between the C band and K sub-bands, is consistent with optically-thin emission of a jet's knot from VLA--1.

%_____________________________________________________________
%-----------------------------------------------------------------------------------------------------------

\subsection{Example\,3: a thermal jet near to a HCH\,{\sc ii} region}\label{ex3}

In Fig.\,\ref{fig3}, we present an example of thermal jet near to a hyper-compact (HC) H\,{\sc ii} region, towards the
H$_2$O maser site G240.32$+$0.07. In panel\,A, the radio continuum emission is dominated by two components, labeled VLA--1
and VLA--2, which are separated by about 4300\,au along the northeast-southwest direction. The peak brightness of component
VLA--1 is more than 10 times higher than that of VLA--2 at each frequency, and both components have a size comparable to that
of the beam (Table\,\ref{contab}). Because of the limited \textit{uv}-coverage, we performed self calibration at each band, in order
to improve on the dynamic range of the maps, which exceeds a factor of 100. At Ku-band, two additional peaks are distinctly
resolved to the north-east and south-west of VLA--1 and VLA--2, respectively, and have been labeled VLA--3 and VLA--4. Component 
VLA--3 is about 50 times less intense than VLA--1, and  VLA--4 is about 6 times less intense than VLA--2  (Table\,\ref{contab}). 
Because of the larger beam size, at the C and K bands VLA--3 and VLA--4 are heavily blended with, respectively, VLA--1 and VLA--2,
and we do not discuss them further\footnote{Flux densities of components VLA--1 and VLA--2 were computed within squared boxes 
centered on each peak. The box size was set so as to minimize the contribution of sources VLA--3 and VLA--4, based on the Ku-band map.}.

In panel\,B, we show that the integrated fluxes of source VLA--1 do not fit a straight line with positive slope, but can be reproduces by a model of 
homogeneous H\,{\sc ii} region with constant density and temperature. We fix the electron temperature of the H\,{\sc ii} region to $10^4$\,K,
taking into account the lower limit to the brightness temperature of 3000\,K at Ku band (optically thin regime). The continuum spectrum in
panel\,B is produced by a number of Lyman photons (N$_{Ly}$)  of 3\,$\times$\,10$^{46}$\,s$^{-1}$ and Str\"{o}mgren radius (R$_S$) of
870\,au, corresponding to an angular size of $0\farcs370$ at a distance of 4.72\,kpc (i.e., the beam size). The H\,{\sc ii} region has an electron
density (n$_e$) of 1.2\,$\times$\,10$^{5}$\,cm$^{-3}$ and emission measure (EM) of 1.2\,$\times$\,10$^{8}$\,pc\,cm$^{-6}$. The radio
continuum emission is strong enough to allow obtaining individual spectral index maps of the C and Ku bands separately (panels~D and~E). These maps
show that the spectral index values within the C and Ku bands are in excellent agreement with the continuum spectrum of the H\,{\sc ii} region,
confirming that the spectral slope is inverted between 4 and 18\,GHz  from positive to negative values. The number of Lyman photons, required
to excite the HCH\,{\sc ii} region, corresponds to that emitted by a zero-age-main-sequence star of spectral type between B1--B0.5
\citep[e.g.,][]{Thompson1984}. This star dominates the bolometric luminosity of the region of 10$^{3.9}$\,L$_{\odot}$ (Table\,1 of Paper\,I).

%We require the H\,{\sc ii} region to have the angular size of the beam at the C and K bands ($0\farcs370$), corresponding to
%a Str\"{o}mgren radius (R$_S$) of  870\,au at a distance of 4.72\,kpc. The continuum spectrum in panel\,B is produced by a
% number of Lyman photons (N$_{Ly}$)  of 3\,$\times$\,10$^{46}$\,s$^{-1}$, and the H\,{\sc ii} region has an electron density
% (n$_e$) of 1.2\,$\times$\,10$^{5}$\,cm$^{-3}$ and emission measure (EM) of 1.2\,$\times$\,10$^{8}$\,pc\,cm$^{-6}$.

In panel\,C, we show that the integrated fluxes of source VLA--2 fit a straight line with angular coefficient ($\alpha$) of 0.84; this 
averaged spectral index is constrained within a small uncertainty of \,$\pm$\,0.03. The radio continuum emission from VLA--2 is
consistent with thermal bremsstrahlung from a radio jet, which is elongated in the southeast-northwest direction. On the other hand,
the spectral index maps of the C and Ku bands encompass a wide range of (positive) spectral index values. At C band (panel~D),
$\alpha$ ranges between 1.0 and 1.3; this range is consistent with a single value within 1\,$\sigma$. At Ku band (panel~E), $\alpha$
ranges between 0 and 1.5; this range exceeds the largest uncertainty on $\alpha$ ($\pm$\,0.2) and increases regularly from the
southeast to the northwest.

According to eq.\,15 of \citet{Reynolds1986}, changes of $\alpha$ with position might be due to changes of the flow geometry 
($\epsilon$), and irregular changes of the temperature (q$_T$) and optical depth ($q_{\tau}$) across the flow. Practical examples
are, respectively, a wide-angle wind which collimates at large radii from the star, a sudden drop of temperature away from the
heating source, and the optical depth which increases towards the receding (red-shifted) lobe of the outflow and decreases towards
the approaching (blue-shifted) lobe.  The first two examples would produce, on the $\alpha$-map, a symmetric gradient with respect
to the central source, whereas the third example would produce a continuous change of $\alpha$ across the flow, from the red-shifted
lobe (higher $\alpha$) to the blue-shifted lobe (lower $\alpha$).

Assuming that the source exciting VLA--2 is at the center of the radio continuum emission, a change in the optical depth, between the
red-shifted jet lobe (to the northwest) and the blue-shifted jet lobe (to the southeast), would explain the regular gradient of $\alpha$.
This interpretation is consistent with the geometry of the molecular outflow detected in the region by \citet{Qiu2009}. In Appendix\,\ref{spvar}, 
we generally comment on the variations observed in the spectral index maps.

%_____________________________________________________________
%                                              FIGURES  N. 5
%-----------------------------------------------------------------------------------------------------------

\onlfig{
\begin{figure*}
%\sidecaption
\centering
\includegraphics [angle= 0, scale= 0.95]{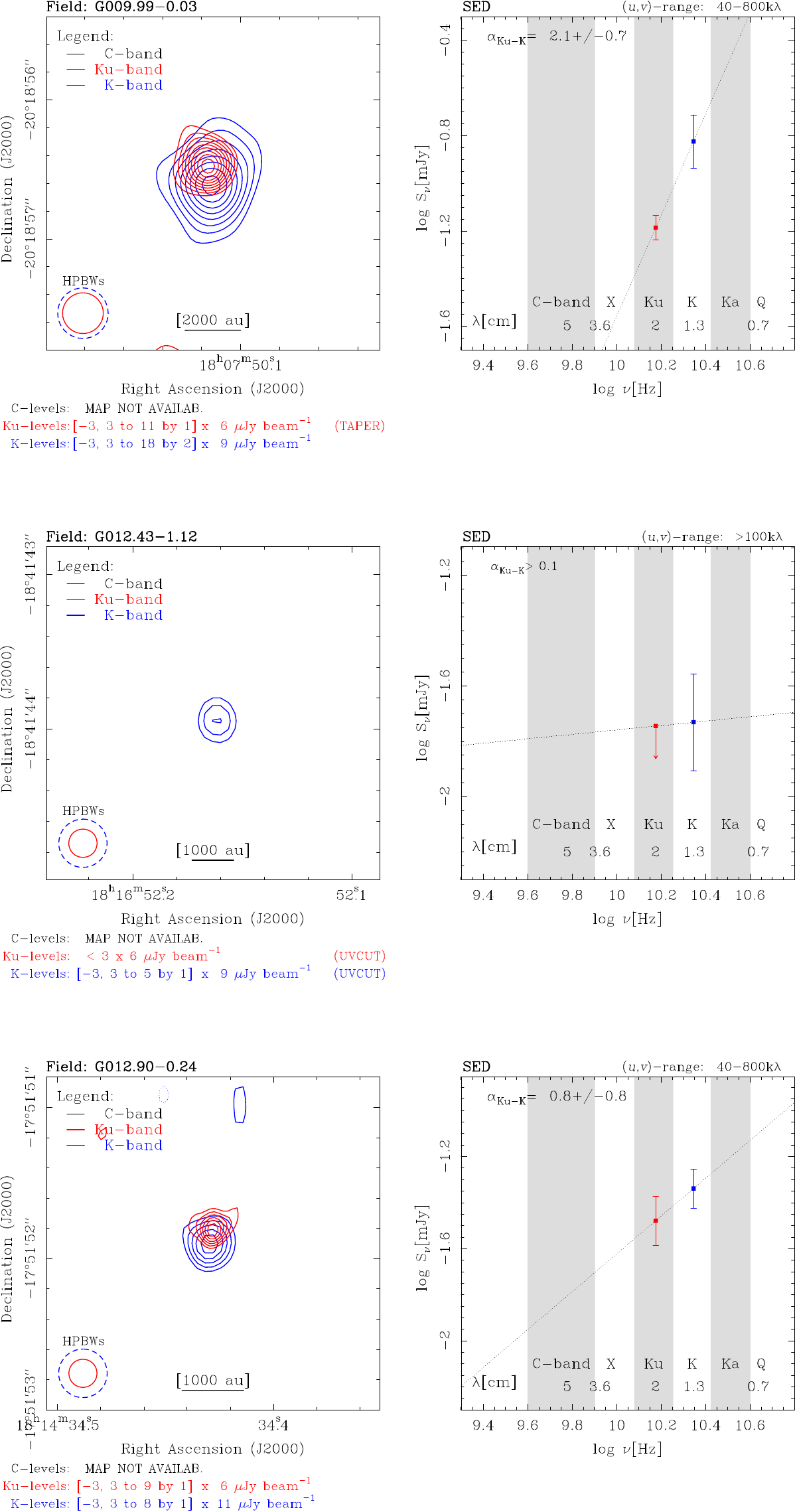}
\caption{For each source in Table\,\ref{targets}, analysis of the VLA radio continuum emission associated with the 
H$_2$O maser sites. Labels and symbols as in Figs.\,\ref{fig1}--\ref{fig4}.}\label{fig5}
\end{figure*}
}

\onlfig{
\addtocounter{figure}{-1}
\begin{figure*}
%\sidecaption
\centering
\includegraphics [angle= 0, scale= 0.95]{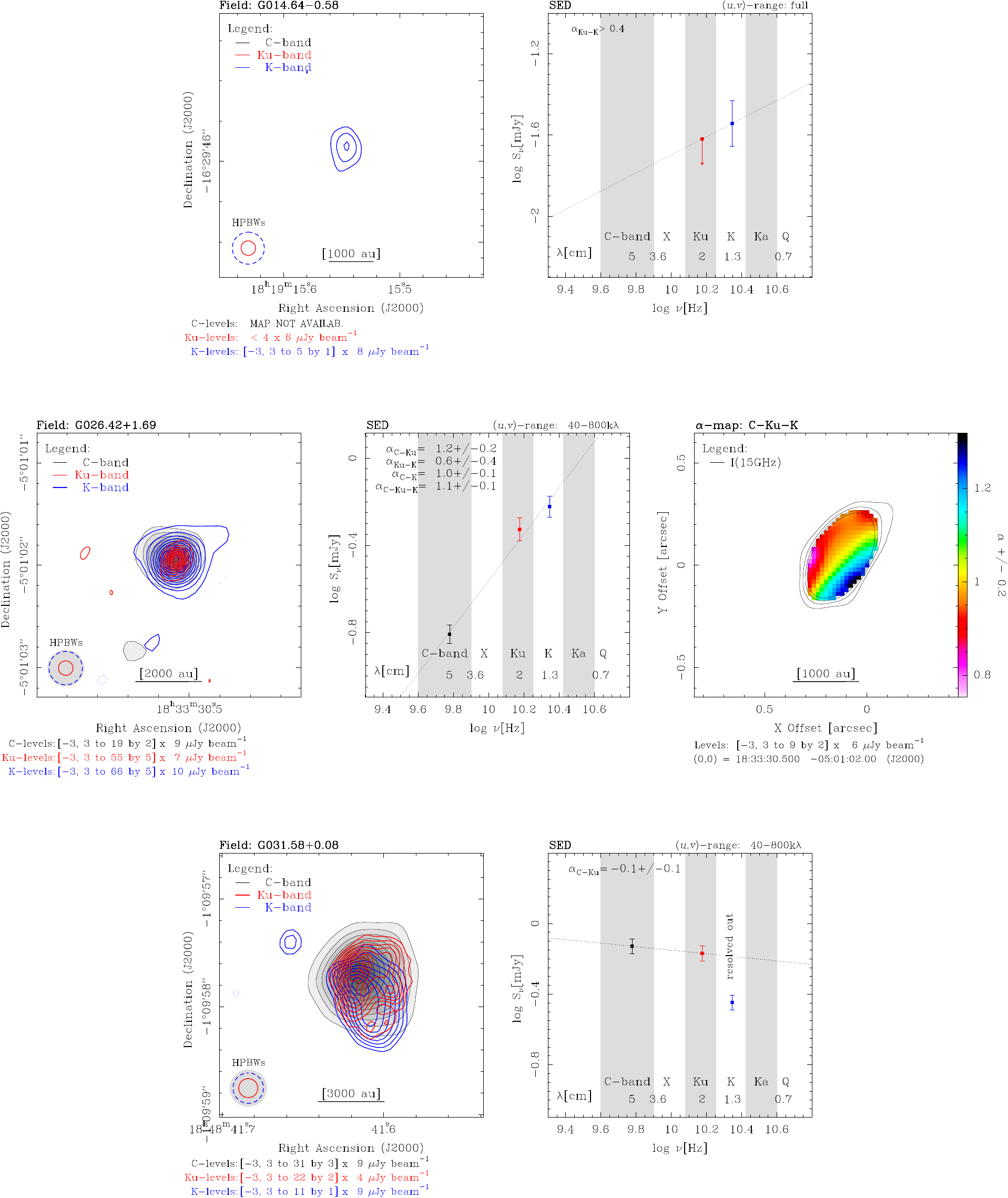}
\caption{(Continued)}
\end{figure*}
}

\onlfig{
\addtocounter{figure}{-2}
\begin{figure*}
%\sidecaption
\centering
\includegraphics [angle= 0, scale= 0.95]{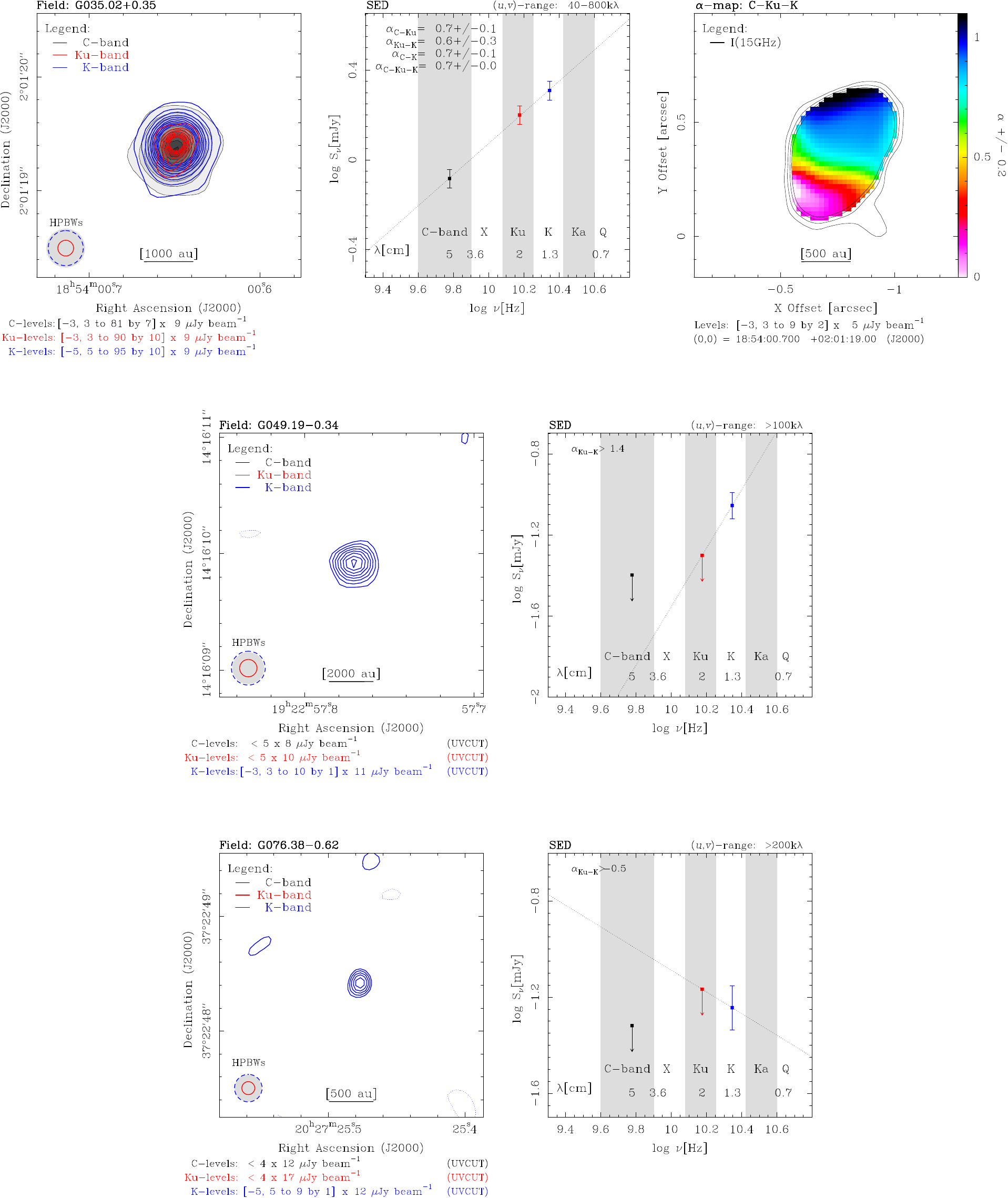}
\caption{(Continued)}
\end{figure*}
}

\onlfig{
\addtocounter{figure}{-3}
\begin{figure*}
%\sidecaption
\centering
\includegraphics [angle= 0, scale= 0.95]{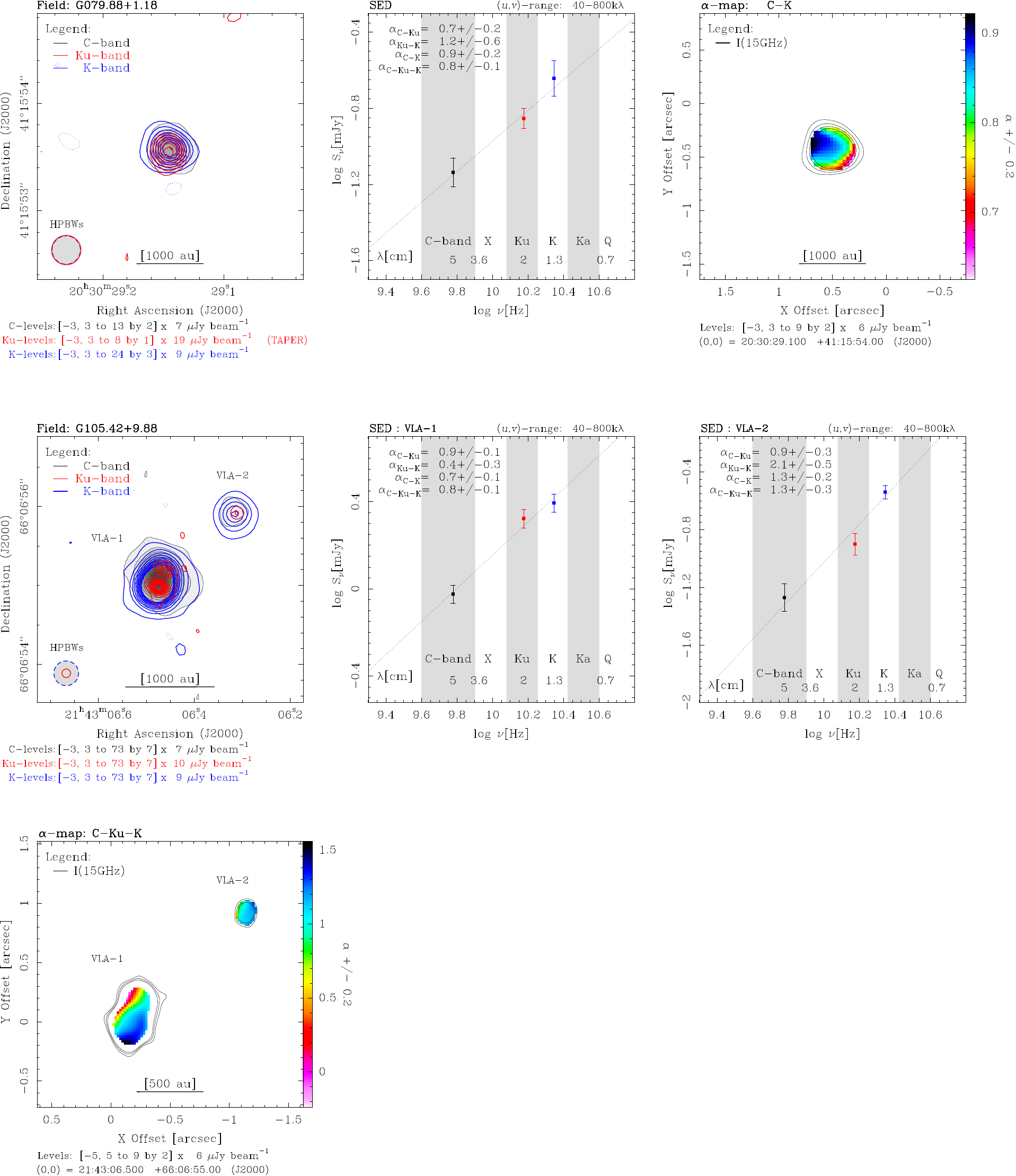}
\caption{(Continued)}
\end{figure*}
}

\onlfig{
\addtocounter{figure}{-4}
\begin{figure*}
%\sidecaption
\centering
\includegraphics [angle= 0, scale= 0.95]{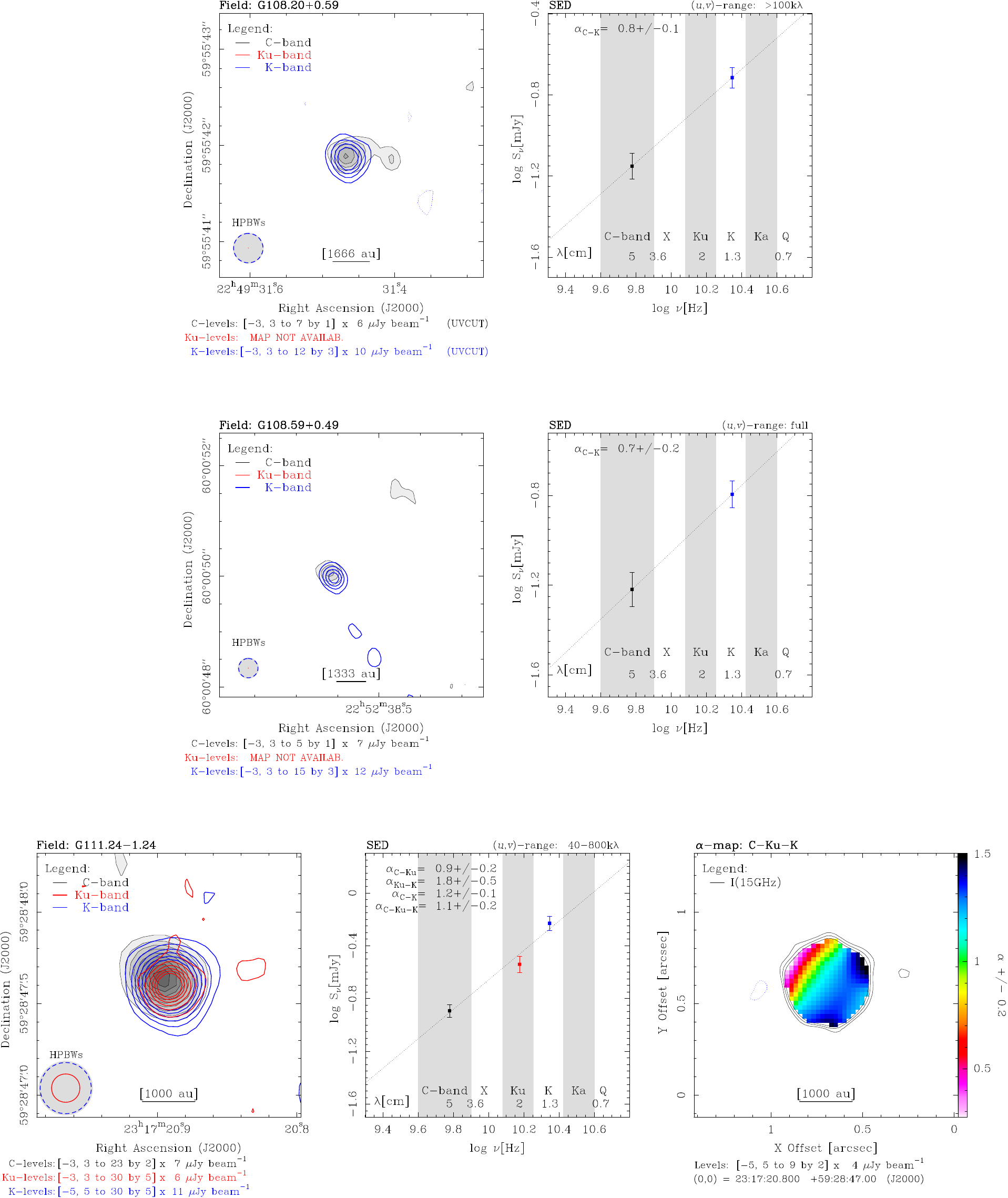}
\caption{(Continued)}
\end{figure*}
}

\onlfig{
\addtocounter{figure}{-5}
\begin{figure*}
%\sidecaption
\centering
\includegraphics [angle= 0, scale= 0.95]{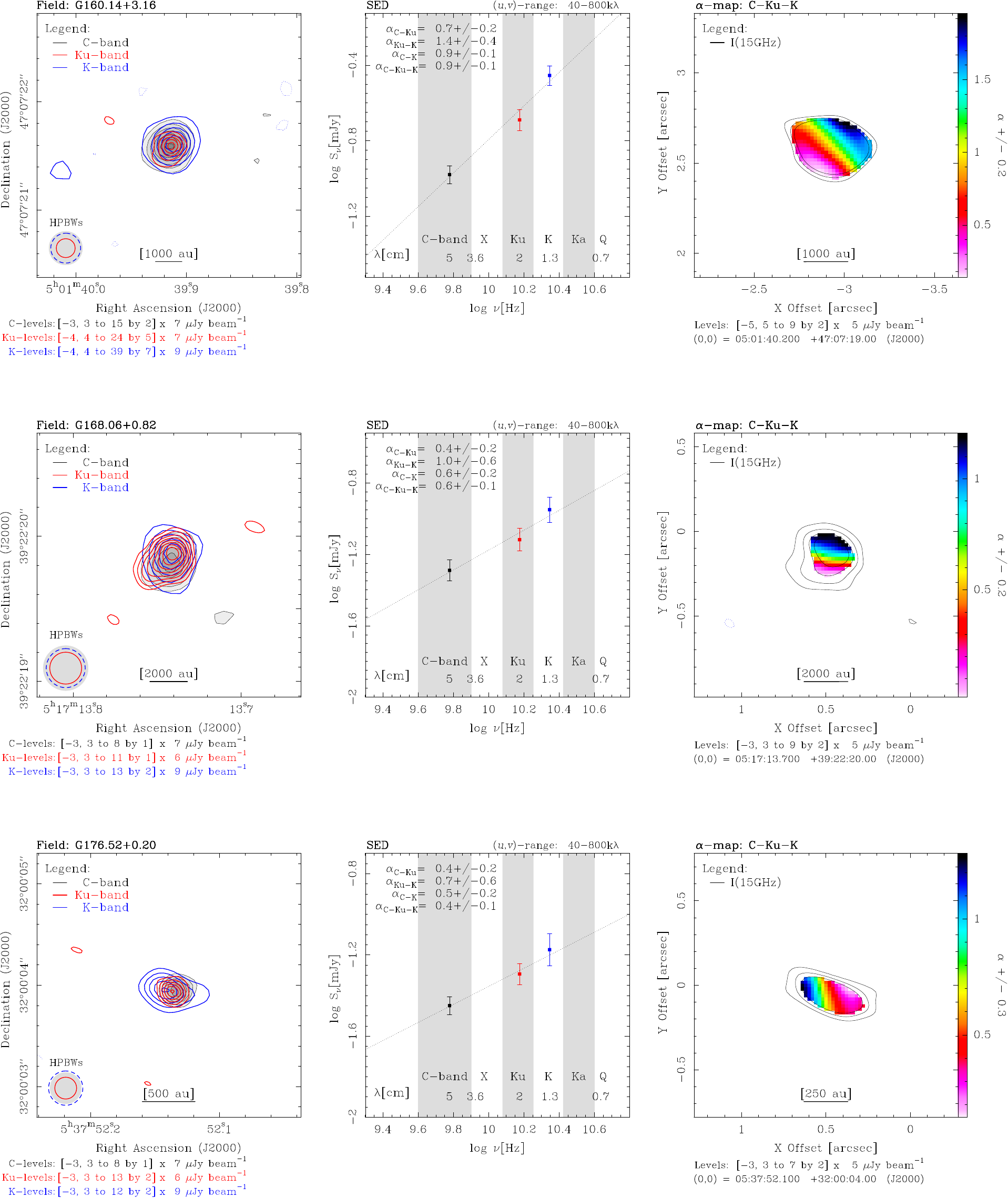}
\caption{(Continued)}
\end{figure*}
}

\onlfig{
\addtocounter{figure}{-6}
\begin{figure*}
%\sidecaption
\centering
\includegraphics [angle= 0, scale= 0.95]{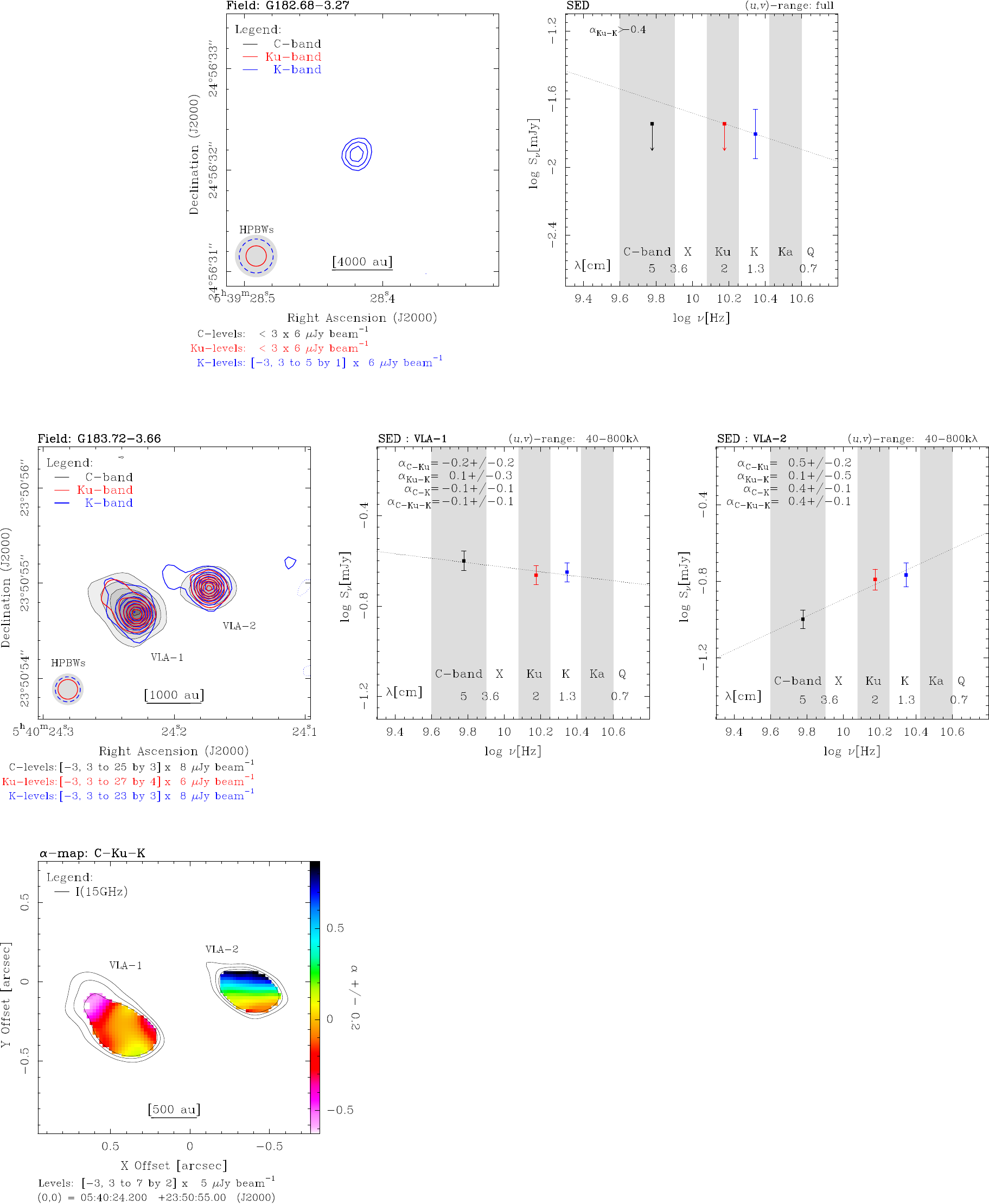}
\caption{(Continued)}
\end{figure*}
}

\onlfig{
\addtocounter{figure}{-7}
\begin{figure*}
%\sidecaption
\centering
\includegraphics [angle= 0, scale= 0.95]{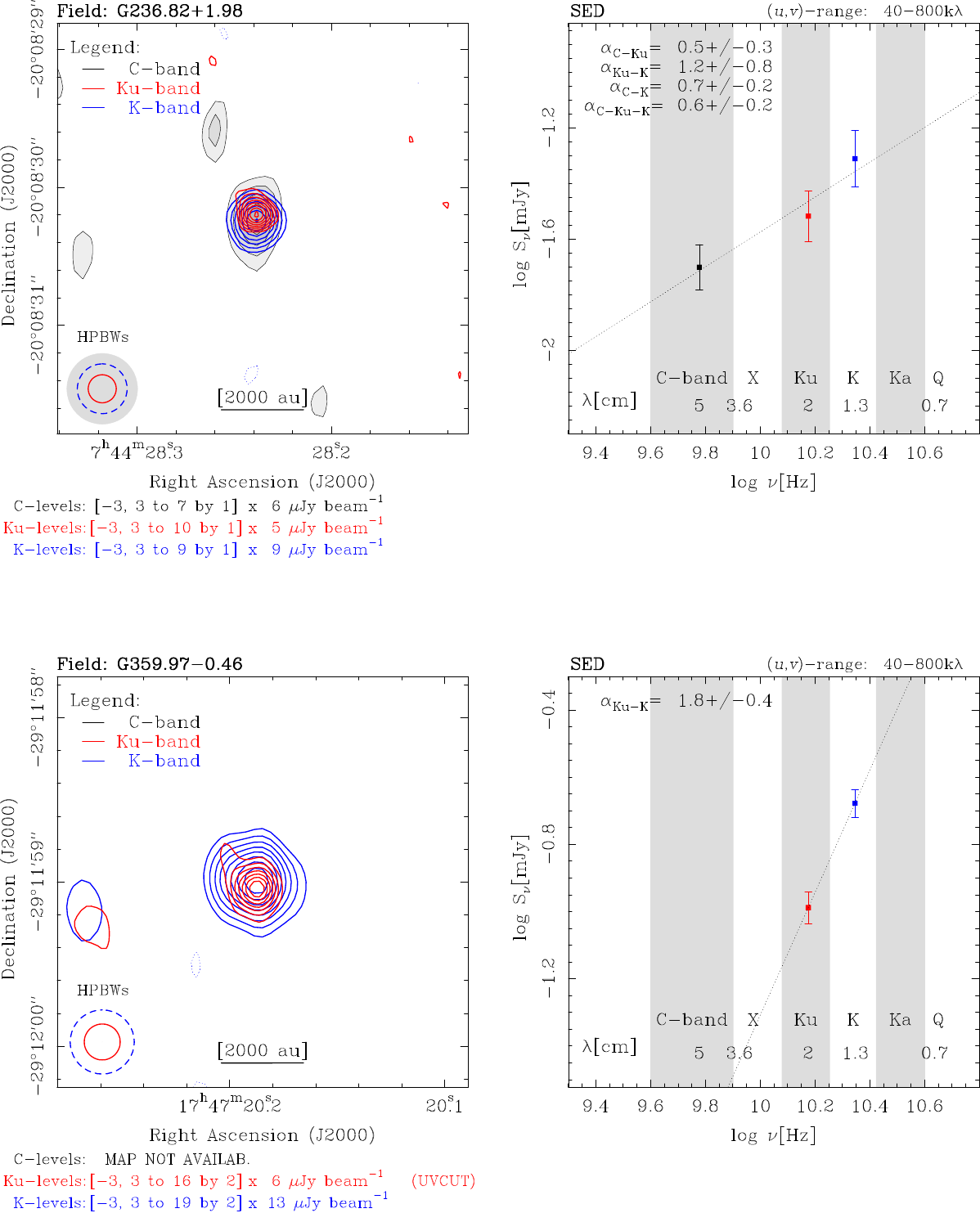}
\caption{(Continued)}
\end{figure*}
}

%_____________________________________________________________
%-----------------------------------------------------------------------------------------------------------

%_____________________________________________________________
%                                              FIGURES  N. 6
%-----------------------------------------------------------------------------------------------------------

\begin{figure*}
\addtocounter{figure}{-7}
%\sidecaption
%\centering
\includegraphics [angle= 0, scale= 1.15]{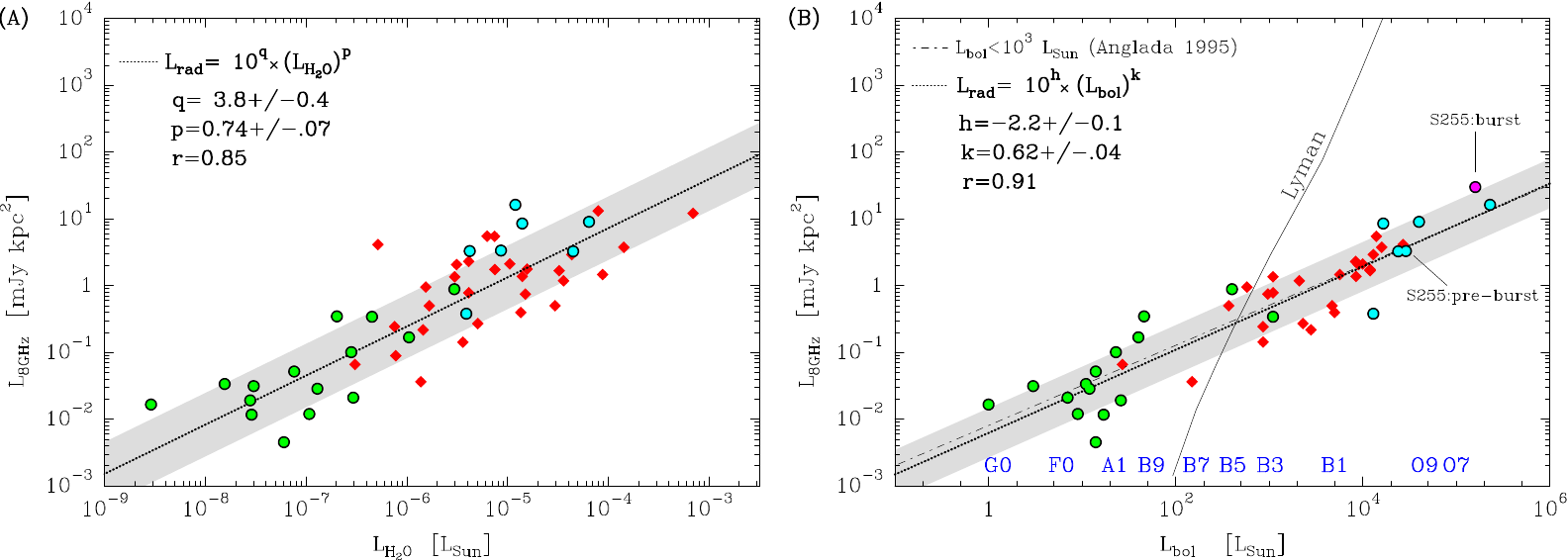}
\caption{Dependence of the radio luminosity of the POETS sample on the H$_2$O maser (left) and bolometric (right) luminosities (Sect.\,\ref{lum}).
Radio continuum luminosities have been normalized to 8\,GHz for comparison with previous works, and a logarithmic scale is used. \textbf{Panel\,A:}  
Red diamonds mark the POETS sample (Table\,\ref{targets}); cyan circles mark additional H$_2$O maser sites associated with luminous, prototypical,
radio thermal jets (Table\,\ref{protojet}); green circles mark low-luminosity H$_2$O maser sites (from \citealt{Furuya2003}). The dotted bold line traces
the $\chi^2$ fit to the sample distribution; the grey shadow marks the dispersion (1\,$\sigma$) about the best fit. The parameters and uncertainties
of the best fit, \textit{q} and \textit{p}, are reported in the upper left, together with the linear correlation coefficient of the distribution ($r$).
\textbf{Panel\,B:} Same symbols as in panel\,A. The parameters of the best fit, \textit{h} and \textit{k}, are to be compared with, e.g.: 
\citet{Anglada2015,Purser2016}. The best fit to the low-mass radio jet luminosites is drawn with a dashed-dotted line (from \citealt{Anglada1995}).
The solid line traces the ionized flux expected from the Lyman continuum of ZAMS stars earlier than B8 (see Sect.\,\ref{lum}). For ZAMS stars,
spectral types, corresponding to a given luminosity, are indicated in blue near the lower axis (from \citealt{Thompson1984}). The pink circle marks
the bolometric and radio luminosities of source S255\,NIRS3 during its recent accretion burst, for comparison \citep{Caratti2017,Cesaroni2018}.}\label{fig6}
\end{figure*}

%_____________________________________________________________
%-----------------------------------------------------------------------------------------------------------

\subsection{Example\,4: a thermal jet with nearby knot -- W33A}\label{ex4}

As a last example, we detail on the case of W33A, for which X-, Ku-, and Q-band data are available from the literature. We targeted this field
at both C and K bands; C-band observations were conducted under program 12B-044 and are described in Paper~I. We have detected two
radio continuum components (Fig.\,\ref{fig4}\,A), labeled VLA--1 and VLA--2, which are separated by about 3300\,au along the northwest-southeast
direction. Source VLA--1 coincides in position with the brightest peak of the main dust condensation MM1 (e.g., Fig.\,1 of \citealt{GalvanMadrid2010}),
and was previously detected at 3.6\,cm, 2\,cm, and 7\,mm by \citet{Rengarajan1996} and \citet{vanderTak2005}. Source VLA--2, which is an order 
of magnitude less bright than VLA--1 at 1\,cm, has never been detected before, and does not coincide with millimeter dust emission (e.g., Fig.\,1 of
\citealt{GalvanMadrid2010}, and Fig.\,1 of \citealt{Maud2017}).  
%The brightness distribution of source VLA--1 is compact under a two-dimensional Gaussian fit, but extended (faint) emission is clearly detected
%towards VLA--2, at both C and K bands (panel\,A). 

In panel\,B, we make use of data from the literature to show that the flux density of source VLA--1 exhibits a steady increase over a wide interval  
of frequencies, from 4 to 43\,GHz. Since we did not observe this field at Ku band, we divided the K-band data in two sub-bands, each one equal to 
half the bandwidth, in order to prove that the spectral slope between the C and K bands is preserved within the K-band itself. The C-band
integrated flux and the two K-band integrated fluxes fit a linear spectral slope of $1.26\pm0.06$, consistent with thermal bremsstrahlung emission
from a stellar wind or jet. The spectral index of source VLA--1 is higher than 1 to a 5\,$\sigma$ confidence: this result points to acceleration or
recombination in the flow, according to \citet{Reynolds1986}. In panel\,B, we also plot the flux densities measured at X (0.79\,mJy) and Ku
(1.86\,mJy) bands from Table\,2 of \citet{Rengarajan1996}, and the Q-band flux density (4.3\,mJy) from Table\,1 of \citet{vanderTak2005}. We
account for a 20\% uncertainty of the absolute flux scale at X, Ku, and Q bands; these observations were conducted on April 1990, February 1984,
and September 2001, respectively.

In panel\,C, we show that the spectral index of radio component VLA--2  ($-0.12\pm0.08$) is that of optically-thin continuum emission. 
For the spectral index analysis, the K-band data were split in two sub-bands as in panel\,B. Radio thermal jet emission is expected to become
totally transparent (S\,$\propto$\,$\nu^{-0.1}$) away from the driving source \citep{Reynolds1986}. Combining this information with the
spatial information of the outflow emission from MM1-main, we interpret source VLA--2 as a knot of the radio jet. The near-infrared and molecular
outflow emission driven by MM1-main is elongated at a position angle (east of north) between $133\degr$ and $145\degr$
\citep{deWit2010,Davies2010,GalvanMadrid2010}. Accordingly, radio components VLA--1 and VLA--2 are aligned along a position angle of
$152\degr$, measured between the K-band peaks. 

% the tangent directions, between the peak of VLA--1 and the 5\,$\sigma$ contour of VLA--2, define a jet aperture of $\pm10\degr$. 

\subsection{L$_{\rm rad}$ versus L$_{\rm H_2O}$ and L$_{\rm bol}$}\label{lum}

In Fig.\,\ref{fig6}, we study the properties of the radio continuum emission for sources associated with H$_2$O masers. We plot the radio luminosity 
of the POETS sample (red diamonds) as a function of the isotropic H$_2$O maser luminosity (Fig.\,\ref{fig6}\,A), and of the bolometric luminosity
of the young stars (Fig.\,\ref{fig6}\,B). For a direct comparison with previous studies, we have interpolated the radio luminosities to 8\,GHz, taking
into account, for each source, the spectral index calculations performed in Sect.\,\ref{sed} (cf. \citealt{Anglada1995,Purser2016,Tanaka2016}). On 
the one hand, the POETS sample was selected so that absolute positions of the H$_2$O masers are known with uncertainties of a few milli-arcseconds,
allowing us to associate the maser emission with the radio  continuum sources without ambiguity. On the other hand, six radio continuum sources are
found in fields with high multiplicity: since the bolometric luminosity\footnote{Bolometric luminosities have been measured with the
\emph{Herschel}\,Hi-GAL fluxes for 60\% of the POETS sample, and probe angular scales of less than $30''$. For the remaining fields, bolometric 
luminosities have been estimated on the basis of archival mid-infrared (WISE and MSX) and far-infrared (IRAS) fluxes (Table\,1 of \citealt{Moscadelli2016}).}
of these sources cannot be established, they are not plotted in Fig.\,\ref{fig6}\,B. These fields are indicated in Table\,\ref{contab},
and add to the two fields excluded in \citet[their Fig.\,13]{Moscadelli2016}. Additional notes are reported in Table\,\ref{contab}.
 
In order to better sample the range of H$_2$O and bolometric luminosities, we complement the analysis by including a number of radio continuum
sources from the literature, which are also associated with H$_2$O masers. We made use of the sample by \citet[green circles]{Furuya2003} to 
cover the lowest H$_2$O maser luminosities (Fig.\,\ref{fig6}\,A), which are also associated with the less-luminous young stars (Fig.\,\ref{fig6}\,B). 
\citeauthor{Furuya2003} monitored the H$_2$O maser emission with the Nobeyama 45\,m telescope towards a sample of young stars, with luminosities
of the order of 10\,L$_{\odot}$ and distances  within a few 100\,pc, typically. Among their sample, we have selected those sources associated with radio
continuum emission, either at 6 or 3.5\,cm (their Table\,4), and whose H$_2$O maser emission was detected at three or more epochs (15 sources
in total). In Fig.\,\ref{fig6}, we plot the averaged H$_2$O maser luminosity from Table\,2 of \citet[corrected for the erratum]{Furuya2003},
and computed the continuum flux at 8\,GHz assuming a standard spectral index of 0.6.  
   
We have further included in our analysis seven sources with bolometric luminosities in excess of 10$^4$\,L$_{\odot}$ (cyan circles in Fig.\,\ref{fig6}).
These sources are associated with prototypical radio jets and H$_2$O masers, whose proper motions have been studied  by our group with the Very
Long Baseline Array (VLBA) in the recent years. Distance measurements, for all sources but HH\,80-81, were obtained by maser trigonometric parallaxes.
In Table\,\ref{protojet}, for each source we have listed the H$_2$O maser luminosity, computed from our previous measurements, and the continuum
luminosity at  8\,GHz, computed from the closest continuum fluxes and spectral index information available in the literature. 

In panel\,A of Fig.\,\ref{fig6}, we show that the radio continuum and H$_2$O maser luminosities are related each other, and the data 
points are distributed along a line with a correlation coefficient ($r$) of 0.85. The dotted bold line draws the best fit to the sample distribution, 
obtained by minimizing a linear relation between the logarithms of the radio and H$_2$O maser luminosities, 
$\log_{10}(\rm L_{8GHz})=p\,{\cdot}\,\log_{10}(\rm L_{H_2O})+q$. Values of $p$ and $q$, with their uncertainties, are reported in the plot.
The lowest and the highest data points in the distribution bias the linear fit to a lower p-value of $0.69\pm0.06$; these two points were excluded 
for the final best fit. 

%We also note that the data point of W33A is an outlier at 3\,$\sigma$ of the average distribution. Quenching of H$_2$O
%maser emission might be related to a later evolutionary stage (e.g., \citealt{Codella1994,Codella2004}). Accordingly, W33A and AFGL\,2591
%(topmost blue circle in Fig.\,\ref{fig6}\,A) exhibit the largest deviations from the linear distribution, and both are associated with bright near-infrared
%emission. 

In panel\,B of Fig.\,\ref{fig6}, we plot the same radio continuum luminosities but with respect to the bolometric luminosities of each source. 
The solid line draws the radio luminosity expected from an optically thin H II region based on the ionizing photons of ZAMS stellar models
by \citet{Thompson1984}. In comparison with panel\,A, the data points also show a strong linear correlation (0.91) but with a less steep slope. 
The dotted bold line draws the best fit to the sample distribution similar to panel\,A, and the best-fit parameters are reported in the plot with 
their uncertainties. We explicitly note that, including or not the topmost (cyan) data point, corresponding to source AFGL\,2591, does not change 
the fitting results. Also, source S255\,NIRS3 has been recently discovered to undergo an accretion burst \citep{Caratti2017}, which triggered 
a progressive increase of radio thermal jet emission \citep{Cesaroni2018}. In panel\,B, we plot the position of source S255\,NIRS3 before and 
after the accretion burst, showing that the radio and bolometric luminosities agree with the linear relation at different times (the data point
of the burst was not used in the fitting).

The power-law dependence between the radio and bolometric luminosities of young stars has been previously explored by other authors
 \citep{Anglada1995,HoareFranco2007,Purser2016,Tanaka2016}. The (extended) sample of H$_2$O maser sources fits a a power-law 
 of $0.62\pm0.04$ (Fig.\,\ref{fig6}\,B), in agreement with the fit by \citet[$k$\,$=$\,$0.64\pm0.04$]{Purser2016}, who compiled 
 a recent sample of thermal jets from the literature (their Fig.\,2). The advantage of the H$_2$O maser sample is twofold: radio continuum 
 sources have been selected on the basis of a common signpost, namely the H$_2$O maser emission, providing a homogeneous sample; 
 the H$_2$O maser sources uniformly sample the range of bolometric luminosites. In particular, the POETS sources are under-luminous with
 respect to the radio luminosity expected from the Lyman continuum of ZAMS stars (for spectral types earlier than B6). This evidence is
 suggestive that the young stars exciting the continuum emission are at an early stage of evolution, prior to the UCH~{\sc ii} region phase, or
 in turn, that their radio continuum emission traces stellar winds and jets in most cases.
 
The rate of Lyman photons from a young star, which sets the maximum radio luminosity from photo-ionized gas, differs greatly at varying
spectral types  (continuous line in Fig.\,\ref{fig6}\,B). For L$_{\rm  bol}$\,$<10^3$\,L$_{\odot}$, young stars are over-luminous with respect
to the photo-ionization limit, and the ionization mechanism is likely related to shocks \citep{Curiel1987,Curiel1989}. It has been suggested that
a single power-law correlation, which fits together the low ($<10^3$\,L$_{\odot}$) and high ($>10^3$\,L$_{\odot}$) bolometric luminosites,
implies that a common mechanism, namely shock-ionization, is responsible for the observed radio emission \citep{Anglada1995,Anglada2015}. 

The correlation we prove in Fig.\,\ref{fig6}\,A, between the radio and H$_2$O maser luminosities, provides an argument in favor of the
shock-ionization as opposed to photo-ionization. Since the H$_2$O maser transition at 22.2\,GHz is inverted by shocks, where stellar winds
and jets impact ambient gas, it follows that, if the radio continuum emission is also produced by shocks, and both are related to the mechanical 
energy of the outflow emission, the radio continuum and H$_2$O maser luminosities should be related\footnote{We note that, even though 
masers might be amplifying a background continuum, we exclude a systematic effect on Fig.\,\ref{fig6}\,A for two reasons:
(i) 22.2\,GHz H$_2$O masers do not necessarily overlap with the 22\,GHz continuum, and (ii) there is no spatial correlation between the 22\,GHz
continuum brightness and the distribution of the brightest H$_2$O masers \citep[e.g.,][]{Moscadelli2016}.}. 
\cite{Felli1992} firstly established a correlation between the H$_2$O maser luminosity and the mechanical energy of molecular outflow, as
measured on parsec scales (see also Sect.\,4.4 of \citealt{Tofani1995}). Here, we revise this correlation with improved confidence, by
directly associating H$_2$O maser spots and radio thermal jets at the smallest scales accessible ($\lesssim0\farcs1$), and with accurate
absolute positions ($\sim 0\farcs01$).

\section{Conclusions}\label{concl}

We report about a multi-frequency VLA survey of radio continuum emission towards a large sample of H$_2$O maser sites (36), entitled: 
the ``Protostellar Outflows at the EarliesT Stages'' (POETS) survey. We extend on the early work by \citet{Tofani1995}, with the idea
of combining the information on maser lines and free-free continuum, to study the engines of molecular outflows. These observations
achieve sensitivities as low as 5\,$\mu$Jy\,beam$^{-1}$ and probe angular scales as small as $0\farcs1$.

Our results can be summarized as follows:  

\begin{enumerate}

\item This is the first systematic survey towards H$_2$O maser sites conducted at high angular resolution and different frequency
bands (C, Ku, and K). At the current sensitivity, the detection rate of 22\,GHz continuum emission, within a few 1000\,au from the
H$_2$O masers, is 100\%.

\item The radio continuum fluxes are well below those expected as due to the Lyman flux of ZAMS stars, suggesting that the young stars
exciting the continuum emission are at an early stage of evolution. Spectral indexes ($\alpha$) between 1 and 7\,cm are generally positive 
and below 1.3; a few sources are associated with optically thin and thick emission at the longer and shorter wavelengths, respectively. 
These results are consistent with the radio continuum emission tracing the ionized gas component of stellar winds and jets. We discuss a
number of scenarios for thermal jet emission, such as in cores with moderate optical depth, surrounded by optically-thin extended emission
and knots (e.g., Sects.\,\ref{ex1}--\ref{ex4}).

\item Radio thermal jets associated with H$_2$O maser emission show a strong correlation ($r$\,$>$\,0.8) with both, the isotropic H$_2$O
maser luminosity, and the bolometric luminosity of the young stars (Fig.\,\ref{fig6}). This seems independent of the spectral type of the young
stars. In turn, these findings support the idea that their radio continuum flux is mainly produced through shock-ionization, given that: (i)
H$_2$O masers also trace shocked gas where stellar winds and jets impact ambient gas, and (ii) photo-ionization is not effective for ZAMS stars
earlier than B5. Alternatively, we can conclude that H$_2$O masers are preferred signposts of radio thermal jets, when the stars are not evolved
and associated with bright radio continuum ($\gg1$\,mJy).

\end{enumerate}

In a following paper in the series, we will study the energetics of (proto)stellar winds and jets, by combining the mass loss rate estimated from
the radio continuum observations with the gas kinematics traced by the H$_2$O maser spots.

%By exploiting the synergy between maser lines and free-free continuum, we eventually want to quantify the outflow
% energetics within a few 1000\,au of very young stars.
% To systematically quantify the energetics (i.e., momentum flux and mechanical luminosity) injected by massive forming
% stars into the ISM at their early stages of evolution, on scales between 10^3 and 10^4 AU.

% suggest that bright  associated with young stars with, before developing an H\,{\sc ii} region, are preferentially
%   associated with thermal (bremsstrahlung) jet emission

\begin{acknowledgements}

Comments from an anonymous referee are gratefully acknowledged.
The authors thanks Sergio Dzib, Carlos Carrasco-Gonz\'alez, Mark Reid, Riccardo Cesaroni, and Roberto Galv\'an-Madrid for fruitful discussion in preparation.
A.S. gratefully acknowledges financial support by the Deutsche Forschungsgemeinschaft (DFG) Priority Program 1573.

\end{acknowledgements}

%---------------------------- REFERENCES ------------------------------

\bibliographystyle{aa}
\bibliography{asanna0506}

%\clearpage
%\begin{appendix}
%\end{appendix}
%\clearpage

%---------------------------- APPENDIX ------------------------------
\begin{appendix}

\section{Variations in the spectral index maps}\label{spvar}

%_____________________________________________________________
%                                              FIGURES  N. A1
%-----------------------------------------------------------------------------------------------------------

\begin{figure*}
%\sidecaption
%\centering
\includegraphics [angle= 0, scale= 0.95]{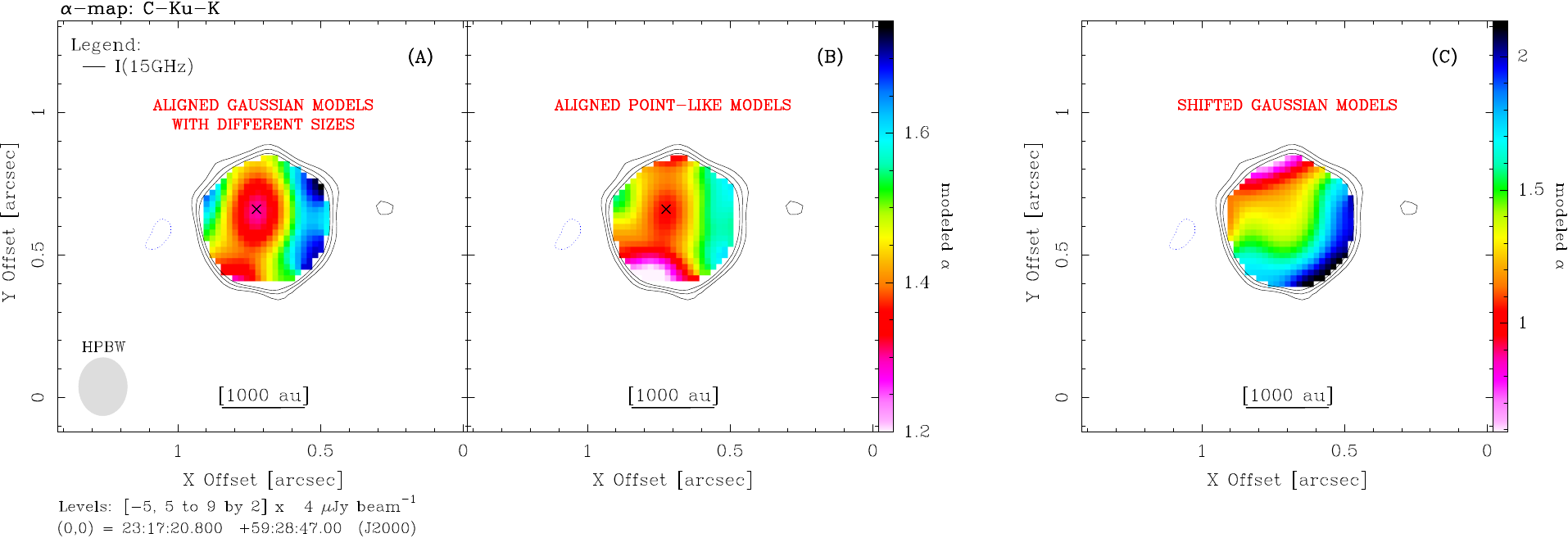}
\caption{Simulated spectral index maps for field G111.24$-$1.24 in Fig.\,\ref{fig5} (Sect.\,\ref{spvar}). Labels and symbols as in Fig.\,\ref{fig1}.
Panels A, B, and C show the spectral index maps (colors) obtained assuming three different source models at C, Ku, and K bands. Black contours are those
of the observed average brightness map for a direct comparison with Fig.\,\ref{fig5}. The cross in panels A and B marks the peak position of the observed 
C-band emission.}\label{figA1}
\end{figure*}

%_____________________________________________________________
%-----------------------------------------------------------------------------------------------------------

In Fig.\,\ref{fig5}, the spectral index maps which we produced by combining observations at different frequency bands show regular gradients 
across the radio continuum emission. Here, we show that these gradients can be produced by small position shifts between the 
radio continuum maps at the different frequencies.  

As an example, we make use of the tasks \emph{simobserve} and \emph{clean} of CASA, in order to simulate the spectral index
map observed for the field G111.24$-$1.24 (Fig.\,\ref{fig5}). The radio continuum emission in G111.24$-$1.24 is compact with
respect to the VLA beam at the different frequencies (column\,4 of Table\,\ref{contab}). We used the CASA toolkit to set the 
source model at the C, Ku, and K bands separately, and assume that the source has a Gaussian brightness distribution, with flux
densities at each frequency from column\,10 of Table\,\ref{contab}.

First, we considered a case where the source size is scaled to a third of the beam at each frequency. We also aligned the Gaussian
peaks at the peak position of the C-band emission (columns\,6 and~7 of Table\,\ref{contab}). We then simulated the
\textit{uv}-datasets at each frequency with the task \emph{simobserve} of CASA,  under the same conditions of our observations
(Table\,\ref{settings}) but setting a long integration time, so that the spectral index maps are not limited by sensitivity. 

In panel\,A of Fig.\,\ref{figA1}, we show the modeled spectral index map cleaned with the same parameters of the observed
map (Sect.\,\ref{amap}). To help the comparison with Fig.\,\ref{fig5}, we also draw the contours of the observed average
brightness map. The modeled spectral index map ranges from 1.2 to 1.7, up to 3\,$\sigma$ from the (average) spectral index
computed with the integrated fluxes (1.1\,$\pm$\,0.2). Close to the C-band peak (cross), and along the beam major axis, spectral
index values are within 1\,$\sigma$ of the average spectral index. For further comparison, we repeated the procedure assuming 
point-like emission at each frequency (panel\,B), but all other parameters being equal. We found that the spectral index distribution
did not change significantly from panels~A to~B.

Finally, in panel\,C of Fig.\,\ref{figA1}, we show a third case where we shifted the peak positions at each frequency according to 
columns\,6 and~7 of Table\,\ref{contab}. The largest offset of $0\farcs07$, between the C- and Ku-band peaks, is within the  
uncertainty of the calibration. The source size was scaled to a third of the beam at each frequency. The modeled spectral index map
spans a larger range of values, from 0.6 to 2.1, and shows a regular gradient from the southwest to the northeast. This gradient
is approximately the same as in the observed map, demonstrating that it is caused by the small offsets between the continuum 
maps at different frequencies. 

This example shows that one has to exert caution when interpreting spectral index gradients in the $\alpha$-maps.

\end{appendix}

%_____________________________________________________________
%                                                    TABLES  # 3  
%-----------------------------------------------------------------------------------------------------------

\onllongtab{
\begin{landscape}
\addtocounter{table}{2}
\begin{longtable}{cccccccccccc}
\caption{Properties of the radio continuum sources.\label{contab}}\\
\hline\hline
\multicolumn{1}{c}{Field} & \multicolumn{1}{c}{Component} & \multicolumn{1}{c}{Band} & \multicolumn{1}{c}{HPBW} & \multicolumn{1}{c}{\emph{rms}} &
\multicolumn{1}{c}{R.A.\,(J2000)}  & \multicolumn{1}{c}{Dec.\,(J2000)} & \multicolumn{1}{c}{I$_{pix}$} & \multicolumn{1}{c}{Size} & \multicolumn{1}{c}{S$_{int}^{\rm SED}$} & \multicolumn{1}{c}{$\alpha$} & \multicolumn{1}{c}{Type} \\
\multicolumn{1}{c}{}  &   &   & \multicolumn{1}{c}{($''$)} & \multicolumn{1}{c}{(mJy\,beam$^{-1}$)}  & \multicolumn{1}{c}{(h\,m\,s)} &
\multicolumn{1}{c}{($^{\circ}$\,$'$\,$''$)} &  \multicolumn{1}{c}{(mJy\,beam$^{-1}$)} & & \multicolumn{1}{c}{(mJy)} &  & \\
(1) & (2) & (3) & (4) & (5) & (6) & (7) & (8) & (9) & (10) & (11) & (12) \\
\hline
\endfirsthead
\caption{continued.}\\
\hline\hline
\multicolumn{1}{c}{Field} & \multicolumn{1}{c}{Component} & \multicolumn{1}{c}{Band} & \multicolumn{1}{c}{HPBW} & \multicolumn{1}{c}{\emph{rms}} &
\multicolumn{1}{c}{R.A.\,(J2000)}  & \multicolumn{1}{c}{Dec.\,(J2000)} & \multicolumn{1}{c}{I$_{pix}$} & \multicolumn{1}{c}{Size} & \multicolumn{1}{c}{S$_{int}^{\rm SED}$} & \multicolumn{1}{c}{$\alpha$} & \multicolumn{1}{c}{Type} \\
\multicolumn{1}{c}{}  &   &   & \multicolumn{1}{c}{($''$)} & \multicolumn{1}{c}{(mJy\,beam$^{-1}$)}  & \multicolumn{1}{c}{(h\,m\,s)} &
\multicolumn{1}{c}{($^{\circ}$\,$'$\,$''$)} &  \multicolumn{1}{c}{(mJy\,beam$^{-1}$)} & & \multicolumn{1}{c}{(mJy)} &  & \\
(1) & (2) & (3) & (4) & (5) & (6) & (7) & (8) & (9) & (10) & (11) & (12) \\
\hline
\endhead
\hline
\endfoot
G009.99$-$0.03\footnote{The 8\,GHz flux of this source was extrapolated assuming an optically-thin spectral index, by analogy with Sect.\,\ref{ex2}.}  & VLA-1\footnote{Tapering of 500\,k$\lambda$ at Ku band.}  & Ku & 0.293 & 0.006 & 18:07:50.117 & --20:18:56.48 & 0.070 & SR     & 0.065   &  $+2.1\pm0.7$ & OP \\
                           &            & K   & 0.364  & 0.009 & 18:07:50.114 & --20:18:56.60 & 0.162 & SR     & 0.150   &  & \\
& & & & & & & & & & & \\
G012.43$-$1.12\tablefootmark{a,b}  & VLA-1\footnote{Maps cleaned with \textit{uv}-cut: $>100$\,k$\lambda$}  & Ku & 0.184 & 0.006 & ... & ... & ... & ...    & $<0.018$   &  $>0.1$ & NA \\
                           &            & K   & 0.313  & 0.009 & 18:16:52.160 & --18:41:43.95 & 0.046 & C  & 0.019   &  & \\
& & & & & & & & & & & \\
G012.90$-$1.12  & VLA-1  & Ku & 0.186  & 0.006 & 18:14:34.427 & --17:51:51.82 & 0.054 & SR  & 0.033  &  $+0.8\pm0.8$ & SW \\
                           &            & K   & 0.321  & 0.011 & 18:14:34.428 & --17:51:51.90 & 0.086 & C   & 0.046   &  & \\
& & & & & & & & & & & \\
G012.91$-$0.26\footnote{C-band data are from exp. 12B-044, and are centered at the frequency of 6.240\,GHz.}  & VLA-1 & C  & 0.358   & 0.013 & 18:14:39.511 & --17:52:00.10 & 0.420  & SR/C  & 0.441  &  $+1.3\pm0.1$ & JET \\
                           &            & K  & 0.325  & 0.011 & 18:14:39.509 & --17:52:00.18 & 2.074  & C  & 1.944/2.668\footnote{Fluxes at the frequencies
of 20.2 and 24.2\,GHz, respectively.}  &  &  \\
                           & VLA-2  & C  & 0.358  & 0.013 & 18:14:39.553 & --17:52:01.24 & 0.223 & SR & 0.197 &  $-0.1\pm0.1$ & \\
                           &            & K   & 0.325 & 0.011 & 18:14:39.551 & --17:52:01.32 & 0.184 & SR  & 0.181/0.157\footnote{Fluxes at the frequencies
of 20.2 and 24.2\,GHz, respectively.}  &  & \\
& & & & & & & & & & & \\
G014.64$-$0.58\tablefootmark{b}  & VLA-1  & Ku & 0.180  & 0.006 & ... & ... & ... & ...    & $<0.024$   &  $>0.4$ & NA \\
                           &            & K   & 0.398  & 0.008 & 18:19:15.545 & --16:29:45.80 & 0.042 & C   & 0.029   &  & \\
& & & & & & & & & & & \\
G026.42$+$1.69  & VLA-1 & C  & 0.356   & 0.009  & 18:33:30.508 & --05:01:01.94 & 0.179  & SR  & 0.156  &  $+1.1\pm0.1$ & SW \\
                           &            & Ku & 0.142  & 0.007  & 18:33:30.508 & --05:01:01.94 & 0.390  & SR   & 0.470  &  & \\
                           &            & K   & 0.327  & 0.010  & 18:33:30.508 & --05:01:01.94 & 0.663  & C    & 0.599   &  & \\
& & & & & & & & & & & \\
G031.58$+$0.08\tablefootmark{b,}\footnote{This source was excluded from Fig.\,\ref{fig6}\,A because it is not a thermal jet.}  & VLA-1 & C   & 0.343  & 0.009  & 18:48:41.616 & --01:09:57.68 & 0.292  & R   & 0.743  &  $-0.1\pm0.1$ &   H\,{\sc ii} \\
                           &            & Ku & 0.177  & 0.005  & 18:48:41.618 & --01:09:57.62 & 0.112  & R   & 0.678  &  & \\
                           &            & K   & 0.282  & 0.010  & 18:48:41.616 & --01:09:57.80 & 0.110  & R   & 0.357\footnote{Resolved out/missing flux}  &  & \\
& & & & & & & & & & & \\
G035.02$+$0.35\tablefootmark{a}  & VLA-1 & C   & 0.338  & 0.009  & 18:54:00.648 &  02:01:19.36 & 0.733  & SR & 0.825  &  $+0.69\pm0.02$ & JET \\
                           &            & Ku & 0.138  & 0.009  & 18:54:00.649 &  02:01:19.42  & 1.140  & R   & 1.584  &  & \\
                           &            & K   & 0.308  & 0.009  & 18:54:00.648 &  02:01:19.42  & 1.872  & C   & 2.036  &  & \\
& & & & & & & & & & & \\
G049.19$-$0.34\tablefootmark{b}  & VLA-1\footnote{Maps cleaned with \textit{uv}-cut: $>100$\,k$\lambda$}  & C  & 0.271   & 0.008  &  ... & ... & ... & ...    & $<0.040$   &  & \\
                           &            & Ku & 0.149  & 0.010  &  ... & ... & ... & ...    & $<0.050$   &  $>1.4$ & NA \\
                           &            & K   & 0.291 & 0.011  & 19:22:57.771 & 14:16:09.94 & 0.115  & C  & 0.088   &  & \\
& & & & & & & & & & & \\
G076.38$-$0.62\tablefootmark{a,b}  & VLA-1\footnote{Maps cleaned with \textit{uv}-cut: $>200$\,k$\lambda$}  & C  & 0.236 & 0.012  &  ... & ... & ... & ...    & $<0.048$   &  & \\
                           &            & Ku & 0.113 & 0.017  &  ... & ... & ... & ...     & $<0.068$   &  $>-0.5$ & NA \\
                           &            & K   & 0.241 & 0.012  & 20:27:25.477 & 37:22:48.40 & 0.119  & C    & 0.057   &  &  \\
& & & & & & & & & & & \\
G079.88$+$1.18  & VLA-1\footnote{Tapering of 500\,k$\lambda$ at Ku band.}  & C  & 0.288 & 0.007 &  20:30:29.143 & 41:15:53.58 & 0.093 & SR & 0.073  &  $+0.8\pm0.1$ & SW \\
                           &            & Ku & 0.272 & 0.019  & 20:30:29.144 & 41:15:53.55  & 0.166  & SR/C  & 0.140  & &  \\
                           &            & K   & 0.265 & 0.009  & 20:30:29.144 & 41:15:53.55  & 0.258  & C       & 0.228   &  & \\
& & & & & & & & & & & \\
G090.21$+$2.32 & VLA-1  & C   & 0.292  & 0.006 & 21:02:22.703 & 50:03:08.30 & 0.130 & R     & 0.153/0.150\footnote{Fluxes at the frequencies
of 5.0 and 7.0\,GHz, respectively.}   &  $-0.1\pm0.1$ & JET \\
                           &            & Ku & 0.166 & 0.006 & 21:02:22.706 & 50.03.08.27 & 0.091 & R      & 0.137   &  &  \\
                           &            & K   & 0.285 & 0.010 & 21:02:22.700 & 50.03.08.27 & 0.354 & SR/C & 0.378/0.475\footnote{Fluxes at the frequencies
of 20.2 and 24.2\,GHz, respectively.}  & $+2.5\pm0.5$ & OP \\
& & & & & & & & & & & \\
G105.42$+$9.88 & VLA-1  & C   & 0.315 & 0.007 & 21:43:06.470 & 66:06:55.06 & 0.579  & SR & 0.944  &   $+0.8\pm0.1$ & JET \\
                           &            & Ku & 0.110 & 0.010 & 21:43:06.474 & 66:06:54.98 & 0.898  &  R  & 2.100  &  & \\
                           &            & K   & 0.313 & 0.009 & 21:43:06.480 & 66:06:55.00 & 1.917  & SR & 2.477  &  & \\
                           & VLA-2  & C   & 0.315 & 0.007 & 21:43:06.322 & 66:06:55.96 & 0.045  & SR & 0.053  &   $+1.3\pm0.3$ & SW  \\
                           &            & Ku & 0.110 & 0.010 & 21:43:06.312 & 66:06:55.92 & 0.112  &  SR & 0.126 &  & \\
                           &            & K   & 0.313 & 0.009 & 21:43:06.312 & 66:06:55.90 & 0.298  &  C  & 0.288 &  & \\
& & & & & & & & & & & \\
G108.20$+$0.59 & VLA-1\footnote{Maps cleaned with \textit{uv}-cut: $>100$\,k$\lambda$}  & C  & 0.317  & 0.006 & 22:49:31.468 & 59:55:41.88 & 0.044 & R   & 0.070  &  $+0.8\pm0.1$ & SW \\
                           &            & K  & 0.306  & 0.010 & 22:49:31.468 & 59:55:41.88 & 0.179 & SR  & 0.193   &  & \\
& & & & & & & & & & & \\
G108.59$+$0.49 & VLA-1  & C  & 0.377  & 0.007 & 22:52:38.620 & 60:00:50.08 & 0.037 & SR & 0.060  &  $+0.7\pm0.2$ & SW \\
                           &            & K  & 0.350  & 0.012 & 22:52:38.612 & 60:00:49.96 & 0.176 & C   & 0.160   &  & \\
& & & & & & & & & & & \\
G111.24$-$1.24  & VLA-1  & C  & 0.296 & 0.007  &  23:17:20.895 & 59:28:47.66 & 0.143 & C   & 0.128   & $+1.1\pm0.2$ & SW\\
                           &            & Ku & 0.155 & 0.006  &  23:17:20.891 & 59:28:47.60 & 0.270 & C   & 0.288   &   & \\
                           &            & K   & 0.279 & 0.011  & 23:17:20.892  & 59:28:47.60 & 0.669 & C   & 0.589   &  &  \\
%                          & VLA-2  & C  & 0.296 & 0.007  &  ... &  ... &  ... &  ...  &  ...    & $+1.1\pm0.2$ \\
%                          &            & Ku & 0.155 & 0.006  &  ... &  ... &  ... &  ...  &  ...    &   \\
%                          &            & K   & 0.279 & 0.011  &  23:17:20.780 & 59:28:47.00 & 0.141 & C   & 0.114  &  \\                           
& & & & & & & & & & & \\
G160.14$+$3.16  & VLA-1 & C   & 0.311 & 0.007  & 05:01:39.918 & 47:07:21.58 & 0.109 & C   & 0.105   & $+0.9\pm0.1$ & SW\\
                           &            & Ku & 0.174 & 0.007  & 05:01:39.914 & 47:07:21.60 & 0.184 & SR  & 0.204   &   & \\
                           &            & K   & 0.276 & 0.009  & 05:01:39.912 & 47:07:21.64 & 0.349 & C   & 0.351    &  &  \\
& & & & & & & & & & & \\
G168.06$+$0.82  & VLA-1 & C   & 0.311 & 0.007  & 05:17:13.741 & 39:22:19.82 & 0.055 & SR  & 0.051  & $+0.6\pm0.1$ & JET \\
                           &            & Ku & 0.220 & 0.006  & 05:17:13.741 & 39:22:19.88 & 0.065 &  R   & 0.076  &  & \\
                           &            & K   & 0.269 & 0.009  & 05:17:13.741 & 39:22:19.88 & 0.116 & SR  & 0.112  &  & \\
& & & & & & & & & & & \\
G176.52$+$0.20  & VLA-1 & C   & 0.310 & 0.007  & 05:37:52.131 & 32:00:03.95 & 0.055 & C  & 0.035  & $+0.4\pm0.1$ & SW \\
                           &            & Ku & 0.215 & 0.006  & 05:37:52.136 & 32:00:03.94 & 0.080 & C  & 0.051  &   & \\
                           &            & K   & 0.338 & 0.009  & 05:37:52.143 & 32:00:03.95 & 0.102 & SR/C  & 0.067  &  & \\
& & & & & & & & & & & \\
G182.68$-$3.27\tablefootmark{b}  & VLA-1 & C   & 0.416 & 0.006  & ... & ... & ... & ...  & $<0.018$  &  & \\
                           &            & Ku & 0.201 & 0.006  & ... & ... & ... & ...  & $<0.018$  & $>-0.4$  & NA \\
                           &            & K   & 0.335 & 0.006  & 05:39:28.418 & 24:56:32.18 & 0.034 & C  & 0.016  & &  \\
& & & & & & & & & & & \\
G183.72$-$3.66\footnote{H$_2$O masers are associated with VLA--1.} & VLA-1  & C   & 0.329 & 0.008  & 05:40:24.231 & 23:50:54.70 & 0.203 & R   & 0.251 & $-0.1\pm0.1$  & JET \\
                           &            & Ku & 0.206 & 0.006  & 05:40:24.228 & 23:50:54.67 & 0.162  & R   & 0.217 &  & \\
                           &            & K   & 0.262 & 0.008  & 05:40:24.231 & 23:50:54.70 & 0.171  & SR & 0.224  & &  \\
                           & VLA-2  & C   & 0.329 & 0.008  & 05:40:24.174 &  23:50:54.94 & 0.121 & C  & 0.100  & $+0.4\pm0.1$ & SW   \\
                           &            & Ku & 0.206 & 0.006  & 05:40:24.174  &  23:50:54.94 & 0.171 & C  & 0.162  &  & \\
                           &            & K   & 0.262 & 0.008  & 05:40:24.174  & 23:50:54.94 & 0.187  & C  & 0.171 &  & \\                           
& & & & & & & & & & & \\
G229.57$+$0.15\footnote{H$_2$O maser luminosities are 0.3\,$10^{-5}$ and 0.4\,$10^{-5}$\,L$_{\odot}$ for VLA--1 and VLA--2, respectively. For the 
purpose of Fig.\,\ref{fig6}\,B, the bolometric luminosity was split between VLA--1 and VLA--2.} & VLA-1  & C   & 0.481 & 0.006 & 07:23:01.845 & --14:41:32.76 & 0.064 & SR & 0.054 &   $+0.68\pm0.02$ & SW \\
                           &            & Ku & 0.194 & 0.005 & 07:23:01.845 & --14:41:32.79 & 0.108 & SR & 0.100  & &  \\
                           &            & K   & 0.412 & 0.007 & 07:23:01.845 & --14:41:32.82 & 0.168 & C  & 0.134  & &  \\
                           & VLA-2  & C   & 0.481 & 0.006 & 07:23:01.804 & --14:41:32.88 & 0.039 & SR & 0.027 &  $+1.2\pm0.1$ & SW \\
                           &            & Ku & 0.194 & 0.005 & 07:23:01.804 & --14:41:32.94 & 0.099 & C  & 0.078  &  & \\
                           &            & K   & 0.412 & 0.007 & 07:23:01.804 & --14:41:32.94 & 0.179 & C  & 0.137  &  & \\
                           & VLA-3  & C   & 0.481 & 0.006 & 07:23:01.825 & --14:41:31.50 & 0.037 & R & 0.047   &  $+0.0\pm0.1$ & JET \\
                           &            & K   & 0.412 & 0.007 & 07:23:01.821 & --14.41.31.80 & 0.032 & R & 0.046    &  & \\
& & & & & & & & & & & \\
G236.82$+$1.98 & VLA-1 & C   & 0.513 & 0.006  & 07:44:28.238 & --20:08:30.24 & 0.040 & C   & 0.020 &  $+0.6\pm0.2$ & SW \\
                           &            & Ku & 0.203 & 0.005  & 07:44:28.238 & --20:08:30.21 & 0.052 & SR & 0.030 &  & \\
                           &            & K   & 0.361 & 0.009  & 07:44:28.238 & --20:08:30.24 & 0.082 & C   & 0.049  &  & \\
& & & & & & & & & & & \\
G240.32$+$0.07\tablefootmark{a,}\footnote{Since H$_2$O masers are associated with VLA--2, only this component was considered in Fig.\,\ref{fig6}\,A.} & VLA-1  & C   & 0.374 & 0.013 & 07:44:52.036 & --24:07:42.16  & 9.611  & R/SR & 12.533 & ... & H\,{\sc ii} \\  
                           &            & Ku & 0.199 & 0.007 & 07:44:52.042 & --24:07:42.19  & 5.871  & R       & 14.306   & &  \\  
                           &            & K   & 0.370 & 0.016 & 07:44:52.031 & --24:07:42.16  & 9.848  & R/SR  & 13.461  & & \\ 
                           & VLA-2  & C   & 0.374 & 0.013 & 07:44:51.975 & --24:07:42.40 & 0.303   & SR  & 0.335  &  $+0.84\pm0.03$ & JET \\
                           &            & Ku & 0.199 & 0.007 & 07:44:51.977 & --24:07.:42.40 & 0.361   & R   & 0.725 &  & \\  
                           &            & K   & 0.370 & 0.016 & 07:44:51.970 & --24:07:42.34  & 0.764   & SR  & 1.005  & & \\ 
                           & VLA-3  & Ku & 0.199 & 0.007 & 07:44:52.071 & --24:07:41.87  & 0.122   & C & 0.137  &   & NA \\
                           & VLA-4  & Ku & 0.199 & 0.007 & 07:44:51.950 & --24:07:42.58 & 0.062    & C  & 0.090 &   & NA \\
& & & & & & & & & & & \\
G359.97$-$0.46\tablefootmark{a,}\footnote{The Ku-band map was cleaned with \textit{uv}-cut: $>40$\,k$\lambda$}$^,$\footnote{The 8\,GHz
flux of this source was extrapolated assuming an optically-thin spectral index, by analogy with Sect.\,\ref{ex2}.}  & VLA-1 & Ku   & 0.219 & 0.006  & 17:47:20.186 & --29:11:59.03 & 0.101 & SR  & 0.103 &  $+1.8\pm0.4$ & OP \\
                           &            & K    & 0.390 & 0.013  & 17:47:20.189 & --29:11:59.00 & 0.248 & C & 0.210 &  & \\
& & & & & & & & & & & \\ %
%\tablefoottext{a}{
\end{longtable}
\tablefoot{Radio continuum sources have been classified as follows (column 12): stellar winds (SW), jets (JET), opaque cores (OP, by analogy with Sect.\,\ref{ex2}),
and H\,{\sc ii} regions ( H\,{\sc ii}). Sources detected at one frequency only have not been classified (NA). 
\tablefoottext{a}{Sources excluded from Fig.\,\ref{fig6}\,B: L$_{\rm bol}$ is uncertain due to high multiplicity in the field.}
\tablefoottext{b}{Since these sources were only detected at K band, and a reliable estimate of the 8\,GHz flux is not available, they
were excluded from the analysis in Fig.\,\ref{fig6}.}
}
\end{landscape}
}
% NOTE: Because of blanded components, squared boxes were used to compute the integrated fluxes.

%-----------------------------------------------------------------------------------------------------------
%-----------------------------------------------------------------------------------------------------------

%_____________________________________________________________
%                                                    TABLES  # 2
%-----------------------------------------------------------------------------------------------------------

\onltab{
\addtocounter{table}{2}
%\begin{landscape}
\begin{table*}
\caption{Bolometric, H$_2$O maser, and radio luminosites of prototypical jets associated with H$_2$O
maser sites (Fig.\,\ref{fig6}; cyan circles).}\label{protojet}
\centering
\begin{tabular}{c c c c c c}

\hline \hline

 Source  &    D         &          L$_{\rm bol}$              &        L$_{\rm H_2O}$             &  L$_{\rm 8\,GHz}$  & References   \\  
             &  (kpc)     &   (10$^{4}$\,L$_{\odot}$)   &   (10$^{-5}$\,L$_{\odot}$)   &    (mJy\,kpc$^2$)   &                    \\  

\hline
 Cepheus\,A~HW2      & 0.7  & 2.4    & 4.50  &  3.24   & 1,2,3,4,5     \\
 HH\,80-81                & 1.7  & 1.7    & 1.41  &  8.47   & 6,3,7,8,9      \\
 IRAS\,20126             & 1.6  & 1.3    & 0.39  &  0.38   & 10,2,11,12   \\
 S255\,NIRS3             & 1.8  & 2.9    & 0.42  &  3.28   & 13,14,15,16  \\
 AFGL\,5142~MM1     & 2.1  & ...      & 0.86  &  3.33   & 17,15,18,19  \\
 AFGL\,2591~VLA--3 & 3.3  & 23     & 1.20  &  16.15 & 20,21,22  \\
 G023.01$-$0.41      & 4.6  & 4.0    & 6.46  &  8.97   & 23,24,25,26,27 \\ 
%& & & & & & & & &  \\
\hline
%& & & & & & & & &  \\
\end{tabular}

\tablefoot{Sources are listed by increasing heliocentric distance. Column\,2: distances to the Sun. 
Column\,3: bolometric luminosities associated with (and dominated by) individual young stars.
For region AFGL\,5142, a reliable estimate of the bolometric luminosity for source MM1 is not available, and this source has not been plotted 
in Fig.\,\ref{fig6}\,B. Column\,4: H$_2$O maser luminosities for Cepheus\,A~HW2 and HH\,80-81 are from \citet{Furuya2003}; all other 
values have been obtained from accurate VLBA measurements that our group have conducted in the recent years. Column\,5: radio continuum 
luminosities at 8\,GHz. Column\,6: references to the papers used for estimating values in columns\,2--5.}
\tablebib{(1)~\citet{Moscadelli2009}; (2)~\citet{DeBuizer2017}; (3)~\citet{Furuya2003}; (4)~\citet{Rodriguez1994}; (5)~\citet{Curiel2006};
(6)~\citet{Rodriguez1980}; (7)~\citet{Marti1993}; (8)~\citet{Carrasco2010}; (9)~\citet{RodriguezKamenetzky2017};
(10)~\citet{Moscadelli2011}; (11)~\citet{Chen2016}; (12)~\citet{Hofner2007};
(13)~\citet{Burns2016}; (14)~\citet{Caratti2017}; (15)~\citet{Goddi2007}; (16)~\citet{Cesaroni2018}; 
(17)~\citet{Burns2017}; (18)~\citet{Goddi2006a};  (19)~\citet{Goddi2011}
(20)~\citet{Rygl2012}; (21)~\citet{Sanna2012};  (22)~\citet{Johnston2013};
(23)~\citet{Brunthaler2009}; (24)~\citet{Sanna2014}; (25)~\citet{Sanna2010b}; (26)~\citet{Sanna2016}; (27)~\citet{Rosero2016}.}
\end{table*}
%\end{landscape}
}
%_____________________________________________________________
%-----------------------------------------------------------------------------------------------------------

\end{document}